%% file: Ktopilnu_HISQonHISQ.tex
\documentclass[preprint,tightenlines,preprintnumbers,aps,eqsecnum,nofootinbib,superscriptaddress,showpacs,floatfix]{revtex4-1}

\usepackage{amsmath,amssymb,amsbsy,amsfonts,bm}
\usepackage{graphicx}
\usepackage{bm}
\usepackage{capt-of}
\usepackage[mathscr]{euscript}
\usepackage{color}
\usepackage{rotating}
\usepackage{slashed}
\usepackage[normalem]{ulem}
\usepackage{float}
\usepackage{tabularx}
\usepackage{hyperref}

\newcolumntype{C}[1]{>{\hsize=#1\hsize\centering\arraybackslash}X}

\newcommand{\order}{{\rm O}}
\newcommand{\ba}{\begin{eqnarray}}
\newcommand{\ea}{\end{eqnarray}}
\newcommand{\be} {\begin{equation}}
\newcommand{\ee} {\end{equation}}

\newcommand{\cpt}{\raise0.4ex\hbox{$\chi$}PT}
\newcommand{\scpt}{S\raise0.4ex\hbox{$\chi$}PT}
\newcommand{\rscpt}{rS\raise0.4ex\hbox{$\chi$}PT}

\def\gtwid{{\,\raise.3ex\hbox{$>$\kern-.75em\lower1ex\hbox{$\sim$}}\,}}
\def\ltwid{{\,\raise.3ex\hbox{$<$\kern-.75em\lower1ex\hbox{$\sim$}}\,}}
\def\rcite#1{Ref.~\cite{#1}}

\def\eqn#1{\label{eq:#1}}

\def\eq#1{Eq.~(\ref{eq:#1})}

\def\figref#1{Fig.~\ref{fig:#1}}

\begin{document}


\preprint{FERMILAB-PUB-18-439-T}

\title{$\vert V_{us}\vert$ from $K_{\ell3}$ decay and four-flavor lattice QCD}

\author{A.~Bazavov}
\affiliation{Department of Computational Mathematics, Science and Engineering,
and Department of Physics and Astronomy,
Michigan State University, East Lansing, Michigan,  48823 USA}

\author{C.~Bernard}
\affiliation{Department of Physics, Washington University, St.~Louis, Missouri, 63130 USA}

\author{C.~DeTar}
\affiliation{Department of Physics and Astronomy, University of Utah, Salt Lake City, Utah, 84112 USA}

\author{Daping~Du}
\affiliation{Department of Physics, Syracuse University, Syracuse, NY, 13244 USA}

\author{A.X.~El-Khadra}
\affiliation{Department of Physics, University of Illinois, Urbana, Illinois, 61801 USA}
\affiliation{Fermi National Accelerator Laboratory, Batavia, Illinois, 60510 USA}

\author{E.D.~Freeland}
\affiliation{School of the Art Institute of Chicago, Chicago, Illinois, 60603 USA}

\author{E.~G\'amiz}
\email{megamiz@ugr.es}
\affiliation{CAFPE and Departamento de F\'{\i}sica Te\'orica y del Cosmos,
Universidad de Granada, 18071 Granada, Spain}

\author{Steven~Gottlieb}
\affiliation{Department of Physics, Indiana University, Bloomington, Indiana, 47405 USA}

\author{U.M.~Heller}
\affiliation{American Physical Society, Ridge, New York, 11961 USA}

\author{J.~Komijani}
\affiliation{School of Physics and Astronomy, University of Glasgow, Glasgow G12~8QQ, United Kingdom}
\affiliation{Physik-Department, Technische Universit\"at M\"unchen, 85748 Garching, Germany}
\affiliation{Institute for Advanced Study, Technische Universit\"at M\"unchen, 85748 Garching, Germany}

\author{A.S.~Kronfeld}
\affiliation{Fermi National Accelerator Laboratory, Batavia, Illinois, 60510 USA}
\affiliation{Institute for Advanced Study, Technische Universit\"at M\"unchen, 85748 Garching, Germany}

\author{J.~Laiho}
\affiliation{Department of Physics, Syracuse University, Syracuse, NY, 13244 USA}

\author{P.B.~Mackenzie}
\affiliation{Fermi National Accelerator Laboratory, Batavia, Illinois, 60510 USA}

\author{E.T.~Neil}
\affiliation{Department of Physics, University of Colorado, Boulder, Colorado, 80309 USA}
\affiliation{RIKEN-BNL Research Center, Brookhaven National Laboratory, Upton, New York, 11973 USA}

\author{T.~Primer}
\affiliation{Department of Physics, University of Arizona, Tucson, Arizona, 85721 USA}

\author{J.N.~Simone}
\affiliation{Fermi National Accelerator Laboratory, Batavia, Illinois, 60510 USA}

\author{R.~Sugar}
\affiliation{Department of Physics, University of California, Santa Barbara, California, 93106 USA}

\author{D.~Toussaint}
\affiliation{Department of Physics, University of Arizona, Tucson, Arizona, 85721 USA}

\author{R.S.~Van~de~Water}
\affiliation{Fermi National Accelerator Laboratory, Batavia, Illinois, 60510 USA}

\collaboration{Fermilab Lattice and MILC Collaborations}
\noaffiliation

\date{\today}

\begin{abstract}

Using HISQ $N_f=2+1+1$ MILC ensembles with five different values of the lattice spacing, including four ensembles with physical
quark masses, we perform the most precise computation to date of the $K\to\pi\ell\nu$ vector form factor at zero momentum
transfer, $f_+^{K^0\pi^-}(0)=0.9696(15)_\text{stat}(12)_\text{syst}$.
This is the first calculation that includes the dominant finite-volume effects, as calculated in chiral perturbation theory at
next-to-leading order.
Our result for the form factor provides a direct determination of the Cabibbo-Kobayashi-Maskawa matrix element
$|V_{us}|=0.22333(44)_{f_+(0)}(42)_\text{exp}$, with a theory error that is, for the first time, at the same level as the
experimental error.
The uncertainty of the semileptonic determination is now similar to that from leptonic decays and the ratio $f_{K^+}/f_{\pi^+}$,
which uses $|V_{ud}|$ as input.
Our value of $|V_{us}|$ is in tension at the 2--$2.6\sigma$ level both with the determinations from leptonic decays and with the
unitarity of the CKM matrix.
In the test of CKM unitarity in the first row, the current limiting factor is the error in $|V_{ud}|$, although a recent
determination of the nucleus-independent radiative corrections to superallowed nuclear $\beta$ decays could reduce the $|V_{ud}|^2$
uncertainty nearly to that of $|V_{us}|^2$.
Alternative unitarity tests using only kaon decays, for which improvements in the theory and experimental inputs are likely in the
next few years, reveal similar tensions and could be further improved by taking correlations between the theory inputs.
As part of our analysis, we calculated the correction to $f_+^{K\pi}(0)$ due to nonequilibrated topological charge at leading order
in chiral perturbation theory, for both the full-QCD and the partially quenched cases. We also obtain the combination of low-energy constants in the chiral effective Lagrangian
$[C_{12}^r+C_{34}^r-(L_5^r)^2](M_\rho)=(2.92\pm0.31)\cdot10^{-6}$.

\end{abstract}

\maketitle

\input intro

\input simulation

\input corrections

\input chiral

\input systematic

\input results

\input phenomenology

\acknowledgments

We thank Matthew Moulson for useful discussions and Bipasha Chakraborty for participating
in an early stage of this analysis.
We thank Johan Bijnens for making his isospin-breaking NLO partially quenched ChPT and isospin-breaking NNLO full QCD codes
available to us, and Bijnens and Johan Relefors for making their FV ChPT code available to us. We thank Zechariah Gelzer for discussions on autocorrelations in the MILC ensembles.

Computations for this work were carried out with resources provided by the USQCD Collaboration, the National Energy Research Scientific Computing Center, the Argonne Leadership Computing Facility, the Blue Waters sustained-petascale computing project, the National Institute for Computational Science, the National Center for Atmospheric Research, and the Texas Advanced Computing Center. USQCD resources are acquired and operated thanks to funding from the Office of Science of the U.S. Department of Energy. The National Energy Research Scientific Computing Center is a DOE Office of Science User Facility supported by the Office of Science of the U.S. Department of Energy under Contract No. DE-AC02-05CH11231. An award of computer time was provided by the Innovative and Novel Computational Impact on Theory and Experiment (INCITE) program. This research used resources of the Argonne Leadership Computing Facility, which is a DOE Office of Science User Facility supported under Contract DE-AC02-06CH11357. The Blue Waters sustained-petascale computing project is supported by the National Science Foundation (awards OCI-0725070 and ACI-1238993) and the State of Illinois. Blue Waters is a joint effort of the University of Illinois at Urbana-Champaign and its National Center for Supercomputing Applications. This work is also part of the ``Lattice QCD on Blue Waters'' and ``High Energy Physics on Blue Waters'' PRAC allocations supported by the National Science Foundation (award numbers 0832315 and 1615006) and used an allocation received under the ``Blue Waters for Illinois faculty'' program. This work used the Extreme Science and Engineering Discovery Environment (XSEDE), which is supported by National Science Foundation grant number ACI-1548562 [161]. Allocations under the Teragrid and XSEDE programs included resources at the National Institute for Computational Sciences (NICS) at the Oak Ridge National Laboratory Computer Center, The Texas Advanced Computing Center and the National Center for Atmospheric Research, all under NSF teragrid allocation TG-MCA93S002. Computer time at the National Center for Atmospheric Research was provided by NSF MRI Grant CNS-0421498, NSF MRI Grant CNS-0420873, NSF MRI Grant CNS-0420985, NSF sponsorship of the National Center for Atmospheric Research, the University of Colorado, and a grant from the IBM Shared University Research (SUR) program.

This work was supported in part by the U.S.\ Department of Energy under grants
No.\ DE-FG02-91ER40628 (C.B.),
No.\ DE-FC02-12ER41879 (C.D.),
No.\ DE-SC0010120 (S.G.),
No.\ DE-FG02-91ER40661 (S.G.),
No.\ DE-FG02-13ER42001 (A.X.K.),
No.\ DE-SC0015655 (A.X.K.),
No.\ DE-SC0010005 (E.T.N.),
No.\ DE-FG02-13ER41976 (D.T.);
by the U.S.\ National Science Foundation under grants
PHY14-14614 and PHY17-19626 (C.D.),
PHY14-17805 (J.L.), and
PHY13-16748 and PHY16-20625 (R.S.);
by the MINECO (Spain) under grants FPA2013-47836-C-1-P and FPA2016-78220-C3-3-P (E.G.);
by the Junta de Andaluc\'{\i}a (Spain) under grant No.\ FQM-101 (E.G.);
by the European Commission (EC) under Grant No.\ PCIG10-GA-2011-303781 (E.G.);
by the U.K.\ Science and Technology Facilities Council (J.K.);
by the German Excellence Initiative and the European Union Seventh Framework Program under grant agreement No.~291763 as well as the
European Union Marie Curie COFUND program (J.K., A.S.K.).
Brookhaven National Laboratory is supported by the United States Department of Energy, Office of Science, Office of High Energy
Physics, under Contract No.\ DE-SC0012704.
Fermilab is operated by Fermi Research Alliance, LLC, under Contract No.\ DE-AC02-07CH11359 with the United States Department of
Energy, Office of Science, Office of High Energy Physics.

\end{document}

%% file: intro.tex
\section{Introduction}

High-precision tests of the unitarity of the Cabibbo-Kobayashi-Maskawa (CKM) matrix, 
as predicted by the Standard Model (SM), are at the forefront of the current flavor 
physics program. Any violation of the unitarity of the CKM matrix, which describes 
flavor-changing interactions, would be evidence of the existence of physics beyond the 
Standard Model (BSM). 

In particular, first-row unitarity, which requires that 
\ba
\Delta_u\equiv\vert V_{ud}\vert^2+\vert V_{us}\vert^2+\vert 
V_{ub}\vert^2 -1
\label{eq:unitarity}
\ea
vanish, is currently the most precisely tested condition. 
Even in the absence of deviations, 
high-precision determinations of the CKM matrix elements involved in the test 
in Eq.~(\ref{eq:unitarity}) put important constraints on the scale of the allowed
new physics~\cite{Gonzalez-Alonso:2016etj}. 

At the current level of precision one can neglect $|V_{ub}|^2$ in Eq.~(\ref{eq:unitarity}). 
Of the other two CKM matrix elements involved,
$|V_{ud}|$ is precisely 
determined from superallowed nuclear $\beta$ 
decays~\cite{Hardy:2018zsb}. It can also be extracted from measurements 
of the neutron lifetime~\cite{Olive:2016xmw} and pion $\beta$ decay~\cite{Pocanic:2003pf}, 
albeit with much larger errors~\cite{PocanicCKM2016}. Improved experimental measurements of
these processes would be interesting because they are theoretically cleaner. 

The best determinations of $|V_{us}|$ are from kaon decays~\cite{Aoki:2016frl}.
The extraction of $|V_{us}|$ from semileptonic kaon ($K_{\ell 3}$) decay requires knowledge of the
form factor at zero momentum transfer, $f_+^{K\pi}(0)$, which is still the largest source of
uncertainty on $|V_{us}|$.
On the experimental side, it is expected that the ongoing and forthcoming experiments
(NA62, OKA, KLOE-2, LHCb and TREK E36) could
reduce the experimental error to $\sim 0.12\%$ within 5 years~\cite{Moulson:2017ive}.
Reducing the theoretical error in the vector form factor calculation is therefore a crucial
task: it is this task that we take up in this paper.

Determinations of $|V_{us}|$ from leptonic kaon and pion decays ($K_{\ell2}$ and $\pi_{\ell2}$),
combined with $f_K/f_\pi$ from lattice QCD, currently have somewhat smaller errors than those
from $K_{\ell3}$. The total error in $|V_{us}|$ from leptonic decays is 0.25\%~\cite{Durr:2016ulb,Carrasco:2014poa,Blum:2014tka,Bazavov:2014wgs,Dowdall:2013rya,Bazavov:2010hj,Durr:2010hr,%
Follana:2007uv,Rosner:2015wva}, while from semileptonic decays it is 0.34\%~\cite{Aoki:2016frl}.  
These leptonic determinations are indirect, however, because they require an external input for 
$|V_{ud}|$, namely Ref.~\cite{Hardy:2018zsb}.
The direct extraction of $|V_{us}|$ from only kaon leptonic decays using $f_K$ as nonperturbative input gives a larger error of
0.46\%.%
\footnote{This error is based on the $N_f=2+1$ FLAG average for $f_K$~\cite{Aoki:2016frl}, which includes only calculations which do
not use $f_\pi$, and thus $|V_{ud}|$, to set the lattice scale~\cite{Blum:2014tka,Bazavov:2010hj,Follana:2007uv}.}

Currently, the value of $\vert V_{us}\vert$ obtained from leptonic kaon decay is $\sim2\sigma$ larger
than the value from semileptonic kaon decay~\cite{Rosner:2015wva}. 
The leptonic decay is mediated by the axial-vector current while the semileptonic decay 
by the vector current. According to the SM, both approaches should give the same $|V_{us}|$, because
the $W$~boson current has a $V-A$ structure. 
Thus, any significant difference should be carefully analyzed. 

In addition, if $|V_{ud}|$ is taken from Ref.~\cite{Hardy:2018zsb}, the leptonic value of
$\vert V_{us}\vert$ is consistent with
unitarity, Eq.~(\ref{eq:unitarity}), but the semileptonic value of $\vert V_{us}\vert$
leads to a $\sim 2\sigma$ disagreement with unitarity.
As we were finishing this work, a paper appeared with a new calculation of the nucleus-independent electroweak radiative corrections
involved in the extraction of $|V_{ud}|$ from superallowed $\beta$ decays with a new approach based on dispersion
relations~\cite{Seng:2018yzq}. 
If this calculation is confirmed, the resulting value of $|V_{ud}|$ would increase the present tension with unitarity. 
Investigating the origin of these tensions and performing even more stringent tests is crucial for 
the internal consistency of the Standard Model.
It is thus necessary to reduce the error on both the experimental and the lattice-QCD inputs entering determinations of $\vert V_{us}\vert$.

In this paper, we focus on semileptonic kaon decay.
The (photon-inclusive) decay rate for $K^0$ can be written~\cite{Cirigliano11}
\ba\label{eq:Kl3def}
\Gamma \left(K^0 \to \pi^- \ell^+ \nu_\ell (\gamma) \right) = \frac{G_F^2m_K^5}{128\pi^3}S_\text{EW}
\left\vert V_{us}f_+^{K^0\pi^-}(0)\right\vert^2 I_{K^0\ell}^{(0)} 
\left(1+\delta_{\rm EM}^{K^0\ell} + \delta_{\rm SU(2)}^{K^0\pi^-}\right),
\ea
where $G_F$ is the Fermi constant as determined by muon decay, $S_\text{EW}=1.0232(3)$ is the universal short-distance electroweak
correction~\cite{Sirlin:1977sv, Sirlin:1981ie,Marciano:1993sh},%
\footnote{This value of $S_\text{EW}$ is from Ref.~\cite{Marciano:1993sh}. 
  We use it because it is the value used for the experimental average in Ref.~\cite{Moulson:2017ive}.} 
and $I_{K^0\ell}^{(0)}$ is a phase-space integral which depends on the shape of the $f_{+,0}^{K^0\pi^+} (q^2)$ form factors given in
Eq.~(\ref{eq:formfac}) below.
The long-distance electromagnetic corrections are parametrized by $\delta_\text{EM}^{K^0\ell}$.
The strong isospin-breaking parameter $\delta_\text{SU(2)}^{K\pi}$ is defined as a correction with respect to the $K^0$ decay:
\ba\label{eq:SU2def}
\delta_\text{SU(2)}^{K\pi} = \left(\frac{f_+^{K\pi}(0)}{f_+^{K^0\pi^-}(0)}\right)^2-1,
\ea
so that $\delta_\text{SU(2)}^{K^0\pi^-} \equiv 0$.
The $K^+$ decay rate, $\Gamma \left(K^+ \to \pi^0 \ell^+ \nu_\ell (\gamma) \right)$, can be obtained by multiplying the
right-hand side of Eq.~(\ref{eq:Kl3def}) with the Clebsch-Gordan coefficient $C^2_{K^+} = 1/2$ and replacing $I_{K^0\ell}^{(0)}$,
$\delta_{\rm EM}^{K^0\ell}$, and $\delta_{\rm SU(2)}^{K^0\pi^-}$ with the analogous $I_{K^+\ell}^{(0)}$,
$\delta_{\rm EM}^{K^+\ell}$, and $\delta_{\rm SU(2)}^{K^+\pi^0}$.
The long-distance electromagnetic corrections, which are mode dependent, were calculated to $\order(e^2p^2)$ in
Ref.~\cite{Cirigliano:2008wn} and are incorporated into the experimental average for $\vert V_{us}\vert f_{+}^{K^0\pi^-}(0)$, adding
a $0.11\%$ uncertainty to the experimental errors.

The input needed from lattice QCD in Eq.~(\ref{eq:Kl3def}) is the vector form 
factor at zero momentum transfer, $f^{K^0\pi^-}_+(q^2=0)$, defined by 
\ba \label{eq:formfac}
\langle \pi^+ \vert V^\mu \vert K^0\rangle  &=& f_+^{K^0\pi^-}(q^2) \left[p_K^\mu+ p_\pi^\mu\right]
+f_-^{K^0\pi^-}(q^2) \left[p_K^\mu- p_\pi^\mu\right]\nonumber\\
 &=& f_+^{K^0\pi^-}(q^2) \left[p_K^\mu
+ p_\pi^\mu - \frac{m_K^2-m_\pi^2}{q^2}q^\mu\right]+ f_0^{K^0\pi^-}(q^2)
\frac{m_K^2-m_\pi^2}{q^2}q^\mu.
\ea
where $V^\mu=\bar s\gamma^\mu u$ and $q\equiv p_K-p_\pi$. 

The most precise value for $f^{K^0\pi^-}_+(q^2=0)$ to date is provided by the $N_f=2+1+1$ 
Fermilab Lattice/MILC calculation in Refs.~\cite{Bazavov:2013maa,Gamiz:2013xxa}, 
$f_+^{K^0\pi^-}(0)=0.9704(\pm0.33\%)$. More recent lattice-QCD calculations 
by the RBC/UKQCD ($N_f=2+1$)~\cite{Boyle:2015hfa} and ETMC 
($N_f=2+1+1$)~\cite{Carrasco:2016kpy} collaborations agree very well with the 
Fermilab Lattice/MILC central value but with larger errors. 
Earlier $N_f=2+1$ calculations with unphysically heavy pions by 
the Fermilab Lattice/MILC~\cite{Bazavov:2012cd} and 
RBC/UKQCD~\cite{Boyle:2013gsa,Boyle:2010bh} collaborations,
as well as the more recent JLQCD calculation in Ref.~\cite{Aoki:2017spo},
yielded smaller values for $f_+^{K\pi}(0)$, but with larger errors.
With the exception of the earliest calculation \cite{Boyle:2010bh}, these $N_f=2+1$
results are compatible with the newer  $N_f=2+1+1$ ones. For comparison, 
the average of the relevant experimental input~\cite{Moulson:2017ive}, $|V_{us}|f_+^{K^0\pi^-}(0)=0.21654(41)$,
has a $0.19\%$ error. This average includes the strong isospin 
and electromagnetic corrections in Eq.~(\ref{eq:Kl3def}) for each decay mode. 

In this work we reduce the main sources of uncertainty in our previous calculation of $f^{K^0\pi^-}_+(q^2=0)$ to reach a total error
of $0.19\%$, obtaining the most precise calculation to date, and matching the current experimental 
uncertainty, for the first time.
The main improvements over our previous calculation~\cite{Bazavov:2013maa,Gamiz:2013xxa} are increased statistics in some key
ensembles, the addition of a new (smaller) lattice spacing, and the correction of finite-volume effects at next-to-leading order
(NLO) in chiral perturbation theory (ChPT).
Preliminary results were presented in Refs.~\cite{Primer:2014eea,Gamiz:2016bpm}.

There are other ways to determine $|V_{us}|$.
Semileptonic hyperon decays unfortunately lack sufficiently precise knowledge of the SU(3)-breaking
corrections, which precludes a competitive determination. 
A conservative estimate of such effects yields an uncertainty of $\sim2\%$~\cite{Mateu:2005wi}.
Inclusive hadronic $\tau$ decays have, in the past, yielded values of $|V_{us}|$ smaller than the
semileptonic kaon determination and, thus, were in even more disagreement with unitarity~\cite{Gamiz:2004ar}.
A more recent analysis~\cite{Hudspith:2017vew} uses lattice QCD to compute dimension-larger-than-4
condensates and, more importantly, employs a dispersive technique to obtain the $K\pi$ branching fractions.
It points to an inclusive-$\tau$ value of $|V_{us}|$ compatible with unitarity~\cite{Hudspith:2017vew},
although it still remains on the low side. An even more promising approach also based on inclusive strange
hadronic $\tau$ decay data is presented in Ref.~\cite{Boyle:2018ilm}.
Its basic ingredients are replacing the operator-product expansion in the relevant sum rules by
lattice hadronic vacuum polarization
functions and optimizing the weight functions to suppress contributions from the high-energy region, where 
the experimental data have poor precision. 
Preliminary results in Ref.~\cite{Boyle:2018ilm} are compatible with both semileptonic and leptonic
determinations (and thus with unitarity), but have larger errors than either. 
Because the total errors on $|V_{us}|$ in Refs.~\cite{Antonelli:2013usa,Hudspith:2017vew,Boyle:2018ilm}
are dominated by experimental uncertainties, it is expected that these determinations will be significantly
improved with new data from the Belle~II experiment \cite{Kou:2018nap}.
Determinations of $|V_{us}|$ from exclusive $\tau$ decays, which use the same nonperturbative inputs as
the leptonic kaon decay determinations, namely $f_{K^\pm}$ and $f_{K^\pm}/f_{\pi^\pm}$, are still in tension
with unitarity~\cite{Amhis:2016xyh}, but this could also change with
future experimental measurements.

This paper is organized as follows.
In Sec.~\ref{sec:simulations} we describe the methodology of the numerical
lattice-QCD simulations and the details of the ensembles, actions, and
correlation functions used.
Section~\ref{sec:corrections} shows how, following the ChPT approaches of Refs.~\cite{Bernard:2017scg} and~\cite{Bernard:2017npd},
one can correct for leading-order finite-volume effects and for the effects of nonequilibrated topological charge
on the ensemble with finest lattice spacing ($a\approx0.042$ fm), respectively.
We discuss the joint chiral interpolation and continuum extrapolation of our data to the physical point in Sec.~\ref{sec:chiral}.
Section~\ref{sec:systematic} analyzes the statistical and systematic uncertainties.
Final results for the form factor $f_+^{K^0\pi^-}$, as well as for the relevant $\order(p^6)$ low energy constants, are presented in
Sec.~\ref{sec:results}.
In Sec.~\ref{sec:pheno}, we use our form factor result to extract a value of $|V_{us}|$ from kaon semileptonic experimental data and
discuss the implications of this value for phenomenology.
Finally, we present our conclusions and the prospects for further improvement in Sec.~\ref{sec:conclusions}.

%% file: simulation.tex
\section{Lattice setup and analysis}

\label{sec:simulations}

The methodology in this work largely follows that of our previous work in 
Refs.~\cite{Bazavov:2012cd,Bazavov:2013maa,Gamiz:2013xxa}. The approach,
pioneered by HPQCD \cite{HPQCD09}, is based on the Ward-Takahashi identity
relating the matrix elements of a vector current to that of the 
corresponding scalar density:
\ba\label{eq:WI}
q^\mu\langle \pi\vert V_\mu^{{\rm lat}} \vert K\rangle Z_V=(m_s-m_u)
\langle \pi\vert S^{{\rm lat}} \vert K\rangle Z_S\ ,
\ea
with $Z_V$ and $Z_S$ the lattice renormalization factors for the vector current 
and scalar density, respectively, where the scalar density is defined
  as the product of the scalar current and the quark masses $(m_s-m_u)$.
Working with staggered fermions, and choosing
$V_\mu^{\rm lat}$ to be the partially conserved, taste singlet, vector current,
and $S^{\rm lat}$ to be its divergence, we have $Z_V=Z_S=1$.  Thus, $S^{\rm lat}$ is a local,
taste-singlet density, with the same flavor content as the vector current, 
$S=\bar s u$. With the above identity and the definition of the form 
factors in Eq.~(\ref{eq:formfac}), one can extract the scalar form factor 
$f_0(q^2)$ at any value of the momentum transfer $q^2$ by using
\ba\label{eq:WIresult}
f_0^{K\pi}(q^2) = \frac{m_s-m_u}{m_K^2-m_\pi^2}\langle \pi\vert S
\vert K\rangle_{q^2}.
\ea
In addition, a kinematic constraint requires $f_+^{K\pi}(0)=f_0^{K\pi}(0)$, so this relation can be 
employed to calculate $f_+^{K\pi}(0)$ from 3-point correlation functions with a scalar
insertion. As already discussed in our first work~\cite{Bazavov:2012cd}, the use of 
a local scalar density instead of a vector current has two main advantages: avoiding the
use of a renormalization factor and avoiding the use of noisier correlation functions with
either a nonlocal vector current or external non-Goldstone
mesons~\cite{HPQCD09,Koponen:2012di}. 

\subsection{Lattice actions, parameters, and correlation functions}

\begin{figure}[tbp]
\centering
\includegraphics[width=0.65\textwidth]{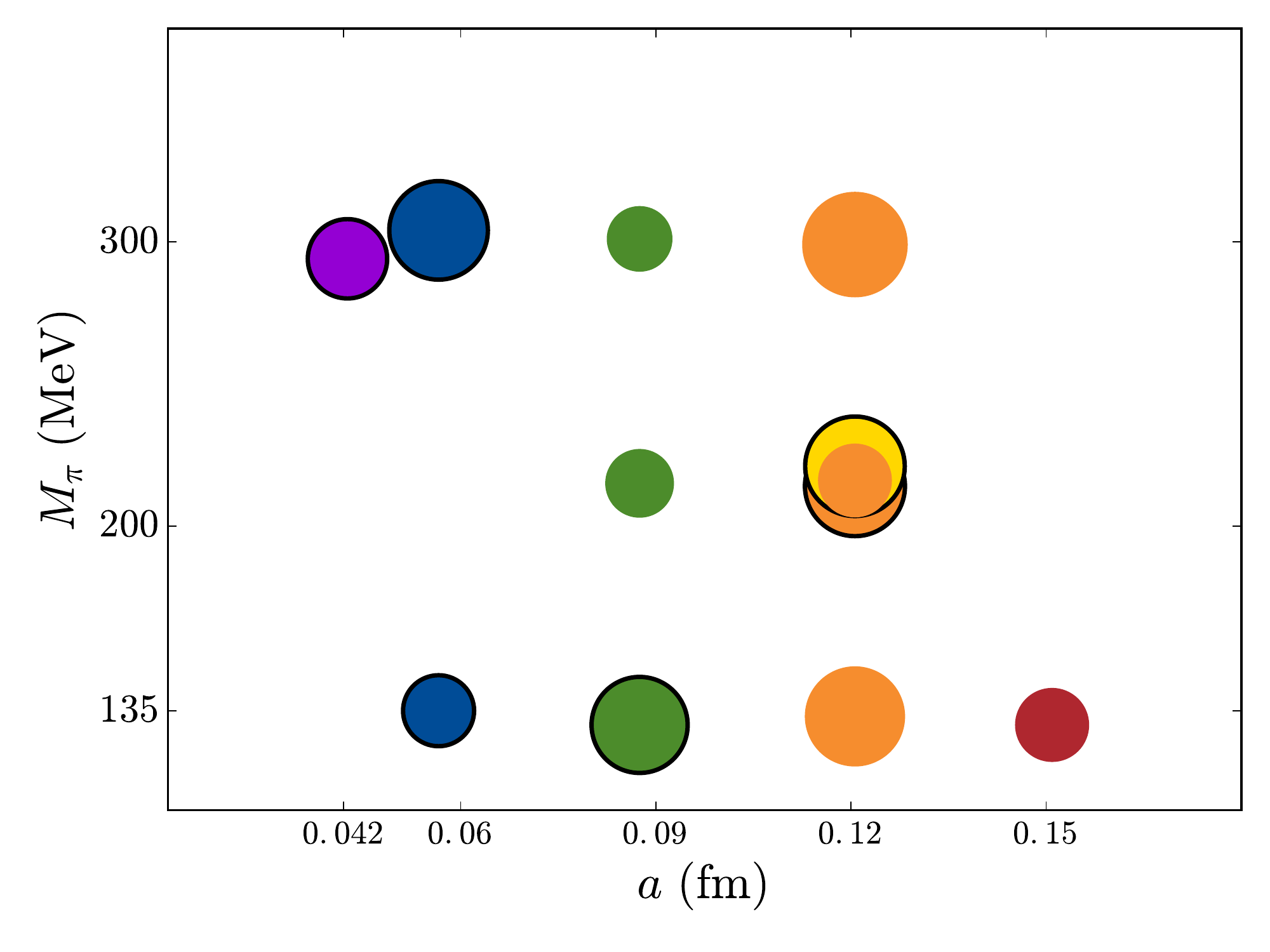}
\caption{Gauge-field ensembles analyzed in this work (parameters of these gauge-field
  ensembles are listed in Table~\ref{tab:ensembles}). The area of each disk 
  is proportional to the statistical sample size $N_{{\rm conf}}\times N_{{\rm src}}$. Ensembles
  on which we have increased the statistics or we have added since our earlier work in
  Ref.~\cite{Gamiz:2013xxa} are indicated with black outlines. The three disks with
  $a\approx 0.12~{{\rm fm}}$ and $m_\pi \approx 200$ MeV correspond to the three ensembles
  with  $m_l/m_s^{{\rm sea}}=0.1$ and different volumes (smaller to larger from 
  top to bottom) in Table~\ref{tab:ensembles}. The yellow disk (smallest volume) is not
  included in the final analysis but used only to
  study finite-volume effects. 
\label{fig:ensembles} }
\end{figure}

We perform our calculation on the highly-improved staggered quark (HISQ) $N_f=2+1+1$ MILC configurations~\cite{Bazavov:2010ru,HISQensembles,Bazavov:2017lyh}
with sea quarks simulated with the HISQ action~\cite{Hisqaction}. 
We also employ the HISQ action for the valence quarks. 
We have already seen in our previous work that 
the use of the HISQ action greatly reduces discretization 
effects~\cite{Bazavov:2013maa,Gamiz:2013xxa}. 
The charm-quark and strange-quark masses on the $N_f=2+1+1$ MILC configurations are 
always tuned to values close to the physical ones, while the light-quark masses 
vary between $0.2m_s$ and $m_s/27$, with the latter approximately the physical value. 
In this work, we include data generated at five different values of the lattice spacing 
down to $a\approx 0.042$~fm, with sea pion masses ranging from $319$ to $134$~MeV. 
Table~\ref{tab:ensembles} lists
the key parameters of the ensembles analyzed here and the correlation
functions calculated on them.  The ensembles include four
with physical quark masses and $a\approx0.15, 0.12, 0.09, 0.06$~fm. 
Ensembles that are new since our analysis in 
Refs.~\cite{Bazavov:2013maa,Gamiz:2013xxa} are marked with a dagger
in the last column; those where we have increased the statistics are marked with an asterisk. 
Table~\ref{tab:ensembles} also lists the pseudoscalar-taste (physical) pion mass $m_{\pi,P}$ 
and the root-mean-squared pion mass $m_\pi^{{\rm RMS}}$ for each ensemble. The
difference is a measure of the dominant discretization effects, which arise from 
taste-changing interactions. As expected, they 
decrease rapidly as the lattice spacing is reduced. 
The data included in this analysis are graphically depicted in
Fig.~\ref{fig:ensembles}.

\begin{table}[thb]
\centering
\caption{
  Parameters of the $N_f=2+1+1$ gauge-field ensembles used in this work, and details
  of the correlation functions generated.
  A dagger at the end of a row indicates an ensemble that is new since our work in 
Ref.~\cite{Bazavov:2013maa}; an asterisk indicates that the statistics have been increased. 
$N_{\rm conf}$ is the number of configurations included in the analysis,
$N_{{\rm src}}$ the number of time sources used on each configuration, and $L$
and $L_t$ the spatial and temporal sizes of the lattice, respectively. 
The column labeled $T$ lists the source-sink separations for the three-point functions generated on each ensemble. 
The $m_\pi$ values are in MeV, with $m_{\pi,P}$ the Goldstone
(pseudoscalar taste)
$\pi$ mass, and $m_\pi^\text{RMS}$ the root-mean-squared (over all tastes) $\pi$ mass. 
The ensemble with $a\approx 0.12~{\rm fm}$, $m_l/m_s=0.1$ and $m_{\pi,P} L=3.2$  
is used solely for the study of finite-volume effects.}
\label{tab:ensembles} 
\vspace{0.5em}
\begin{tabular}{cccc|r@{$\times$}lcccccc}
\hline\hline
$\approx a({\rm fm})$ & $m_l/m_s^\text{sea}$ & $m_{\pi,P} L$ & $L^3\times L_t$ & 
$N_\text{conf}$&$N_\text{src}$ & $T$ & $am_s^\text{sea}$ & $am_s^\text{val}$ & $m_{\pi,P}$ & 
$m_\pi^\text{RMS}$ & \\
\hline
0.15   & 0.035 & 3.2 & $32^3\times48$ & 1000&4 & 12,13,15,16,17,18 & 0.0647 & 0.06905 & 130 & 314 & \\
\hline
0.12   & 0.2 & 4.5 & $24^3\times64$ & 1053&8 & 15,18,20,21,22 & 0.0509  & 0.0535 & 299 & 364 & \\
& 0.1 & 3.2 & $24^3\times64$ & 1020&8 & 15,18,20,21,22 & 0.0507 &  0.053 & 221 & 303 & $\dagger$ \\
& 0.1 & 4.3 & $32^3\times64$ & 993& 4 & 15,18,20,21,22 &  0.0507 & 0.053 & 216 & 299 & \\
& 0.1 & 5.4 & $40^3\times64$ & 1029& 8 & 15,18,20,21,22 & 0.0507 &  0.053 & 214 & 298 & * \\
& 0.035 & 3.9 & $48^3\times64$ & 945& 8 & 15,18,20,21,22 & 0.0507  & 0.0531 & 133 & 246 & \\
\hline
0.09   & 0.2 & 4.5 & $32^3\times96$ & 773&4 & 23,27,32,33,34 &  0.037 & 0.038 & 301 & 323 & \\
& 0.1 & 4.7 & $48^3\times96$ & 853&4 & 23,27,32,33,34 &   0.0363 & 0.038 & 215 & 221 & \\
& 0.035 & 3.7 & $64^3\times96$ & 950&8 & 23,27,32,33,34 & 0.0363 & 0.0363 &  130 & 176 & * \\
\hline
0.06   & 0.2 & 4.5 & $48^3\times144$ & 1000&8 & 34,41,48,49,50 & 0.024 & 0.024 & 304 & 308 & * \\
& 0.035 & 3.7 & $96^3\times192$ & 692&6 & 31,39,40,48,49 &  0.022 & 0.022 & 135 & 144 & $\dagger$ \\
\hline
0.042 & 0.2 & 
4.3 & $64^3\times192$  & 432&12 & 40,52,53,64,65 & 0.0158 & 0.0158 & 294 & 296 & $\dagger$ \\
\hline\hline
\end{tabular}
\end{table}

The structure of the three-point function with a scalar insertion that we 
generate to access the matrix element in Eq.~(\ref{eq:WIresult}) 
is the same as in our previous work~\cite{Bazavov:2012cd,Bazavov:2013maa,Gamiz:2013xxa}. 
We generate light quarks at a time slice $t_{{\rm src}}$ and extended
strange propagators at a fixed distance $T$ from the source. For each 
configuration we have $N_{{\rm src}}$ time sources placed at 
$t_{{\rm src}}=t_0,\,t_0+L_t/N_{{\rm src}},\,t_0+2L_t/N_{{\rm src}}\dots$, 
where $L_t$ is the temporal length of the lattice. The time $t_0$ varies randomly 
from configuration to configuration in an interval $[0,L_t/N_{{\rm src}}]$ to reduce
autocorrelations. 
Roughly following Ref.~\cite{McNeile:2006bz} we use random-wall sources at the pion
source time $t_{{\rm src}}$. On that spatial time-slice, we choose four stochastic
color-vector fields from a Gaussian distribution, with support
on all three colors, and compute light-quark propagators from each of the four sources.
Between the source  and the kaon sink at time $t_{\rm src}+T$,  we contract the extended
strange propagator with a light propagator to form the scalar density. We then study the
$t$ dependence to isolate the desired matrix element.

The light-quark masses are always the same in the sea and valence sectors, 
while the sea and valence strange-quark masses are slightly different in some of
the ensembles\footnote{At the time the analysis
began, the physical value of $am_s$ on those ensembles had been determined
more accurately than when the ensembles were generated. We used the more
accurate values for the valence strange-quark mass to
be closer to the physical point.}; see Table~\ref{tab:ensembles}. 
Table~\ref{tab:ensembles} also lists 
the number of configurations and time sources on each ensemble. We compute three-point 
functions as described above for 5 or 6 different values of the source-sink 
separation $T$, listed in Table~\ref{tab:ensembles}, which correspond 
to approximately the same physical distances across ensembles.
We include both even and odd values of $T$ to disentangle the effects from oscillating states in the correlation functions.

We simulate directly at zero momentum transfer, $q^2\approx0$, by tuning the
external momentum of the pion using partially twisted boundary conditions. 
In particular, we tune 
\ba\label{twistangle}
|\vec\theta_2|=  \frac{L}{\pi}\,            
\sqrt{\left(\frac{m_K^2+m_\pi^2}{2m_K}\right)^2-m_\pi^2},
\ea
with $\vec\theta_2$ the twist angle of the daughter propagator going from the pion 
to the current. The rest of the 
propagators are generated with periodic boundary conditions, the same as in the 
sea sector. We always have diagonal twist angles, 
$\vec\theta_2=|\vec\theta_2|(1,1,1)/\sqrt{3}$, which turn out 
to give smaller finite-volume effects than twisting in only one 
direction~\cite{Bernard:2017scg}. The values of 
$|\vec\theta_2|$ for each ensemble, as well as the corresponding momentum of the pion, 
are given in Table~\ref{tab:momenta}. 

For each ensemble, we generate zero-momentum two-point $\pi$ and $K$ correlation functions 
and two-point $\pi$ correlation functions with external momentum given by the twist 
angle $\vec\theta_2$, defined in Eq.~(\ref{twistangle}). These correlators  
are given by
\ba\label{eq:2ptcorr}
C_{{\rm 2pt}}^P(\vec{p};t) = \frac{1}{L^3}\sum_{\vec{x}}\sum_{\vec{y}}
\,\langle\Phi_P^{\vec{p}}(\vec{y},t+t_{{\rm src}})\Phi_P^{\vec{p}\,\dagger}
(\vec{x},t_{\rm src})\rangle,
\ea
where the interpolating operator $\Phi_P^{\vec{p}\,\dagger}(\vec{x},t)$ creates a meson
$P=\pi,K$ at time $t$ with momentum $\vec{p}$.  The random wall sources automatically 
implement the sum over $\vec{x}$. 
We also generate three-point correlation functions with the kaon at rest: 
\ba\label{eq:3ptcorr}
C_{{\rm 3pt}}^{K\to\pi}(\vec{p}_\pi,\vec{p}_K\!=\!0;t,t_{{\rm src}},T) =
\frac{1}{L^3}\sum_{\vec{x},\vec{y},\vec{z}}
\,\langle \Phi_K^{\vec{p}_K=0}(\vec{x},t_{{\rm src}}\!+\!T)S(\vec{z},t)\Phi_\pi^{\vec{p}_\pi\,\dagger}
(\vec{y},t_{\rm src})\rangle,
\ea
where the pion recoil momentum $\vec{p}_\pi$ is either equal to zero or to
the values listed in Table~\ref{tab:momenta}. 
The scalar density is a local taste-singlet.

\begin{table}
  \caption{Twisting angles and external momenta injected in the three-point functions. 
    The quark masses $am_l$ and $am_s^\text{sea}$ are the same as in Table~\ref{tab:ensembles}, and $\vec{\theta}_2$ is the twisting angle for the light daughter propagator in the pion, defined in Eq.~(\ref{twistangle}).
    The superscript $P$ in the pion masses refers to the pseudoscalar taste. 
    \label{tab:momenta}}
  \centering
\begin{tabularx}{0.54\textwidth}{ccc|C{0.25}C{0.25}}
\hline\hline
 $\approx a({\rm fm})$ & $m_l/m_s^\text{sea}$ & $m_{\pi,P} L$ & $\vert\vec{\theta}_2\vert$ & 
$\vert a\vec{p}_P\vert$\\ 
\hline
0.15   & 0.035 & 3.2 & 1.80966 & 0.17766 \\ 
\hline
0.12   & 0.2   & 4.5 & 0.84749 & 0.11094 \\ 
       & 0.1   & 3.2 & 0.98192 & 0.12853 \\
       & 0.1   & 4.3 & 1.30923 & 0.12853 \\
       & 0.1   & 5.4 & 1.63653 & 0.12853 \\
       & 0.035 & 3.9 & 2.16464 & 0.14168 \\
\hline
0.09   & 0.2   & 4.5 & 0.82675 & 0.08117 \\
       & 0.1   & 4.7 & 1.45024 & 0.09492 \\
       & 0.035 & 3.7 & 2.08413 & 0.10230 \\
\hline
0.06   & 0.2   & 4.5 & 0.81673 & 0.05345 \\
       & 0.035 & 3.7 & 2.01756 & 0.06602 \\
\hline
0.042 & 0.2 & 4.3 & 0.78006 & 0.03829 \\
\hline\hline
\end{tabularx} 
\end{table}

\subsection{Fit methods and statistical analysis}

The fitting strategy we follow to extract the physical quantities from 
our correlation functions has already been discussed in 
Refs.~\cite{Bazavov:2012cd,Lattice2012}. We fit the two-point 
correlation functions for a pseudoscalar meson $P$ to the expression
\ba\label{eq:2ptfit}
C_{{\rm 2pt}}^{P} (\vec{p}_P;t) & = & \sum_{m=0}^{N_\text{exp}} (-1)^{m(t+1)}(Z_m^P)^2
\left(e^{-E_P^m t}+e^{-E_P^m(L_t-t)}\right),
\ea
using Bayesian techniques. In Eq.~(\ref{eq:2ptfit}), $L_t$ is the temporal size 
of the lattice. The oscillating terms with $(-1)^{m(t+1)}$ do not appear for a zero-momentum 
$\pi$. We fit the three-point correlation functions to
\ba\label{eq:3ptfit}
        C_{{\rm 3pt}}^{K\to \pi} (\vec{p}_\pi,\vec{p}_K;t,T) & = &
\sum_{m,n=0}^{N_\text{exp}^\text{3pt}} (-1)^{m(t+1)} (-1)^{n(T-t+1)}A^{mn}(q^2)Z_m^\pi 
Z_n^K\nonumber\\
&&\times\left(e^{-E_\pi^mt}+e^{-E_\pi^m(L_t-t)}\right)
\left(e^{-E_K^n(T-t)}+e^{-E_K^n(L_t-T+t)}\right),
\ea
where the pion and kaon energies and amplitudes, $E_\pi^n$, $E_K^n$, $Z_n^\pi$ and
$Z_n^K$, are the same as those appearing in the two-point fit functions. 

We first fit the two-point functions one by one on each ensemble and check the stability 
of the ground state masses/energies and amplitudes under the choice of fitting range,  
$t\in[t_\text{min},t_\text{max}]$, and the number of exponentials included in the fit 
function [$N_\text{exp}$ in Eq.~(\ref{eq:2ptfit})].
We always include the same number $N$ of regular and oscillating states in those fits, 
i.e., $N_\text{exp}=2N$. 
To evaluate the relative quality of the fits we use the  
  $\chi^2/dof$ and the $Q$ value defined in Ref.~\cite{Bazavov:2016nty}, a quality of fit statistic adapted for fits with Bayesian priors that is similar to the standard $p$ value~\cite{Olive:2016xmw}. By construction, $Q\in[0, 1]$ with larger $Q$ values indicating
  greater compatibility between the data and fit function given the prior constraints ---see 
 Ref.~\cite{Bazavov:2016nty} for details and explicit formulas. In particular, we disregard any
  fit with $Q<0.1$. We also disregard fits with $\chi^2/dof\lesssim0.05$, since those low $\chi^2/dof$ are generally an indication of a bad identification of the ground state and the corresponding fits tend to be unstable with the variation of $t_{\text min}$, number of exponentials, and/or bootstrap resampling. 
  
We observe that for most 
of the choices of time range and for all ensembles, fits stabilize when including 
2+2 or 3+3 states. From that parameter-scanning procedure, we select an optimal set of 
fit parameters for the two-point functions, with a common $t_\text{min}$ for all the 
functions on the same ensemble. Fixing $N=3$, the chosen $t_\text{min}$ is the smallest
  value for which the ground state parameters for all relevant two-point functions reach a plateau 
  and, in addition, for which the fit results (central values, errors, and quality) are stable under variations of  the number of exponentials and bootstrap resampling. 
For a fixed $t_\text{min}$, $t_\text{max}$ is chosen, in general, as the value for which
  fit results are insensitive to the addition of late-time data for which statistical errors are 
larger.
  
We then use those $[t_\text{min},t_\text{max}]$ ranges to perform a fully
correlated combined Bayesian fit including the two- and three-point
functions needed to extract $f_0^{K\pi}(0)$: $\pi$ two-point correlation
functions with and without momentum, $K$ two-point correlation functions
without momentum, and $N_T$ three-point correlation 
functions with $q^2\approx0$,  where $N_T$ is the number of source-sink separations $T$ included in the combined fit. In general, we include three-point functions only at $N_T = 3$ or 4 different values of $T$ out of the 5 or 6, for which we have data. However, the $T$ values included in the combined fits generally cover most of the available range, corresponding to a physical range of $\approx 0.5 - 1.0$~fm. This allows us to resolve excited states while at the same time including data with good ground state contributions. We find that the resulting fits are stable under variations of time range, number of exponentials, and bootstrap resampling. We also find that adding more $T$ values does not improve the quality (error and stability) of the fits. Table~\ref{tab:3ptfits} lists our parameter choices for the combined three-point function fits. 
  
\begin{figure}[tbp]
\centering
\includegraphics[width=0.9\textwidth]{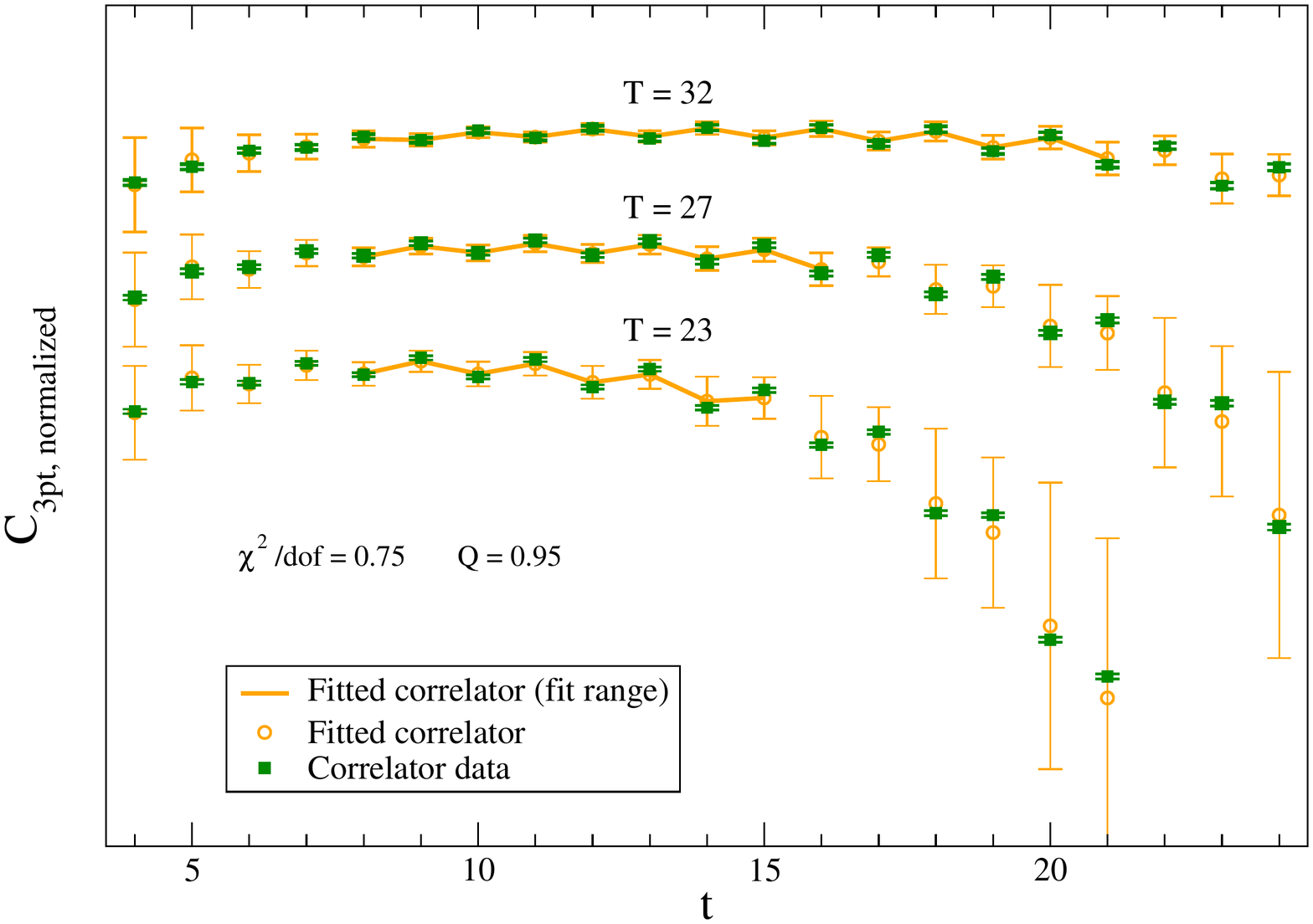}
\caption{
Comparison of data and fit results for the rescaled three-point functions defined in
    Eq.~(\ref{eq:c3ptnorm}) on the $a\approx 0.09$~fm ensemble with physical quark masses. Green
    squares are the data points and orange circles are obtained from the fit posteriors. The fit
    includes the three three-point functions shown, with $T=23,27,32$, in the fit ranges shown
    by the orange lines. Correlators with $T=27,23$ are given a vertical offset, different
    for each $T$, so results for the three correlators do not lie on top of each other. Errors are
    statistical only. 
\label{fig:corr3pt}}
\end{figure}

In general, we use three-point data in the combined fits with 
$t\in[t_\text{min},T-t_\text{min}]$, where $t_\text{min}$ is the value optimized for the two-point 
functions. However, on some ensembles, especially the largest ones, we need to either
shorten the three-point fit range or thin the three-point data in order to obtain 
an acceptable fit, as measured by the $\chi^2/dof$ and $Q$ value.    
A comparison of fit results to data for the $a\approx 0.09$~fm ensemble with
  physical quark masses, one of the most relevant in our analysis, is given in
  Fig.~\ref{fig:corr3pt}. This is a typical case, the comparisons of the fits and data
  on the other ensembles are similar.
  The figure plots the rescaled three-point functions, in which the time-dependent contributions
  of the kaon and pion ground states are removed:
  \ba\label{eq:c3ptnorm}
  C_{3pt,{\rm rescaled}} = \frac{C_{3pt}^{K\to\pi}(\vec{p}_\pi,0;t,T)}{Z_0^\pi Z_0^K
    \left(e^{-E_\pi^0t}+e^{-E_\pi^0(L_t-t)}\right)\left(e^{-E_K^0(T-t)}+e^{-E_K^0(L_t-T+t)}\right)}\,.
  \ea
  In the absence of excited state contributions $C_{3pt,{\rm rescaled}}$ would be time independent. Figure~\ref{fig:corr3pt} shows the comparison of the rescaled correlation functions included in the fit (filled green points) with the results from the fit (open orange circles). We see that the $C_{3pt,{\rm rescaled}}$ exhibit plateaus with a mild oscillation that is more pronounced for the smaller $T$ value, but which can be accounted for almost entirely by the first oscillating state included in the fit.
The agreement between data and the fit is excellent, especially for the time ranges 
  included in the fit, marked by the orange lines. For times closer to the source or the sink, the
  large errors on the orange points indicate a substantial contribution from the excited states
  included in the fit that the data cannot constrain accurately. Nevertheless, the falloff of
  the correlators is well described by the fit functions.

\begin{figure}[tbp]
\centering
\includegraphics[width=0.45\textwidth]{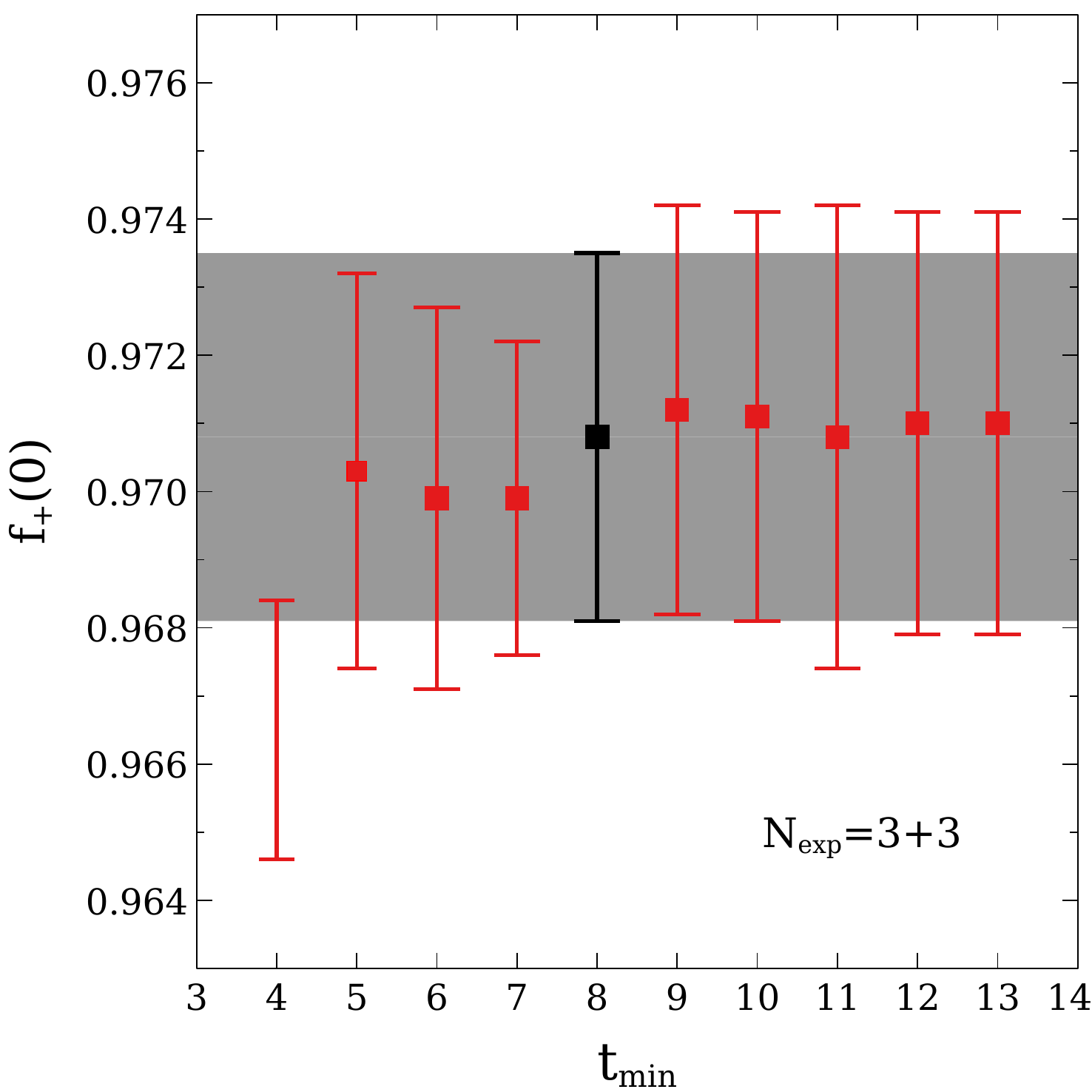}
\hspace*{0.8cm}\includegraphics[width=0.45\textwidth]{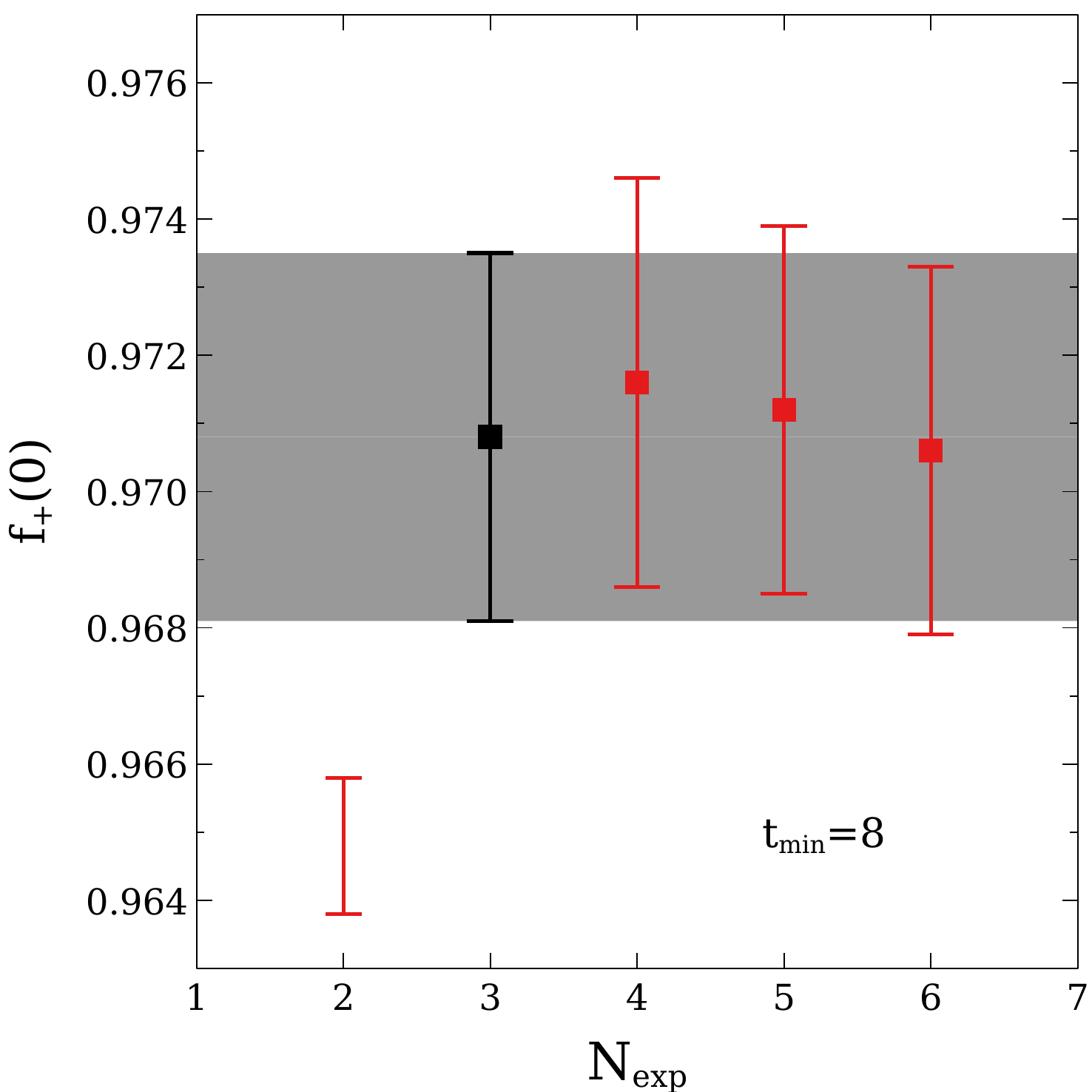}
\caption{
Variation of the fit result for $f_+(0)$ with $t_\text{min}$ (left panel) and $N_\text{exp}$
    (right panel) for the ensemble with $a\approx 0.09$~fm and physical light-quark
    masses. The errors are statistical, generated with a 500-bootstrap distribution.
    The black point on each figure corresponds to our preferred fit with $t_\text{min}=8$ and
    $N_\text{exp}=3+3$. 
\label{fig:stability3pt}}
\end{figure}

From the combined fits, we extract the scalar form factor at zero momentum 
transfer via
\ba\label{eq:f0-extract}
f_0^{K\pi}(0) = 2A^{00}(0)\, \sqrt{E_\pi m_K}\,\frac{m_s-m_l}{m_K^2-m_\pi^2},
\ea
where $A^{00}(0)$ is the ground state three-point parameter in Eq.~(\ref{eq:3ptfit}), 
the meson masses and energies are the values extracted from the combined fits and 
$m_s$ and $m_l$ are the valence strange and light-quark masses simulated.

We check the stability of the combined fit results under the variation of fit ranges, 
number of states, and number and values of source-sink separations included, and choose 
a preferred fit for each ensemble so the shift on the central value with those 
variations is well under the statistical error, and the error is also stable.
Examples of these stability studies are shown in Fig.~\ref{fig:stability3pt}.
  On a few ensembles, the stability tests lead to slight adjustments of our chosen value
  of  $t_\text{min}$, from the $t_\text{min}$ determined in the two-point only fits discussed above. 
  Our final choices of $t_\text{min}$ correspond to very similar physical distances,
  approximately 0.6--0.7~fm, on each ensemble.
  The number of exponentials is always chosen to be 3+3, since also for these combined fits 
  adding more exponentials does not change the fit results and also does not improve fit stability. 
  The parameter values used in the preferred combined fits are listed in Table~\ref{tab:3ptfits}.

\begin{figure}[tbp]
\centering
\includegraphics[width=0.45\textwidth]{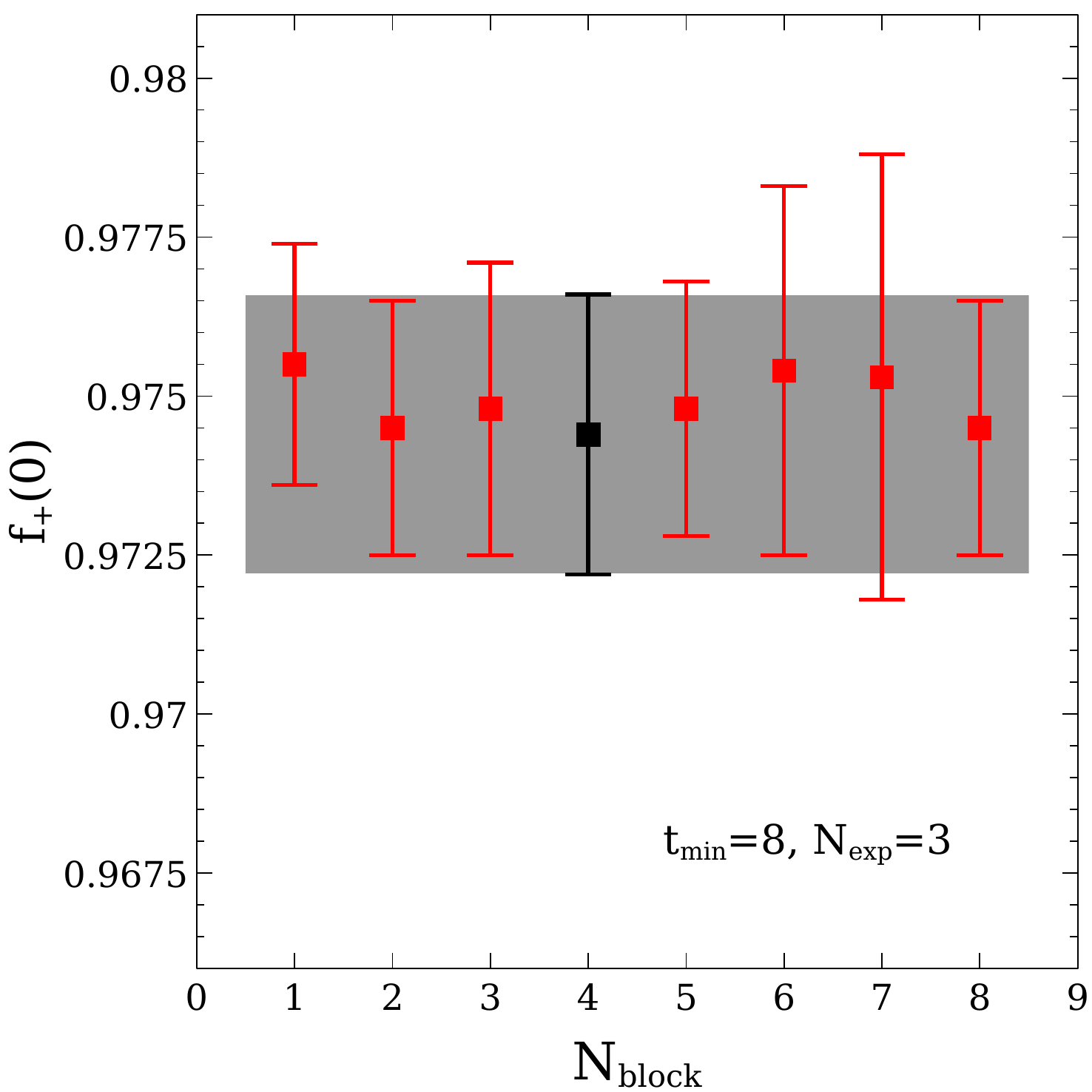}
\caption{
 Variation of $f_+(0)$ with the block size for the ensemble with
    $a\approx 0.15$~fm and physical quark masses. The errors are statistical, generated
    with a 500-bootstrap distribution.
    The black point corresponds to our preferred fit with $N_\text{block}=4$.
\label{fig:block}}
\end{figure}

We study the effect of autocorrelations by blocking the data by increasing numbers of 
successive  configurations and redoing the analysis. 
We do not see evidence of significant autocorrelations on all ensembles, but for
  those where we do see significant changes in central value and error, stability is
  reached with a block size of four. 
For ensembles where we observe significant changes in central value and error for the 
form factor with blocking, those effects stabilize when blocking by four. 
An example for the ensemble with $a\approx0.15$~fm and physical quark masses is given in
  Fig.~\ref{fig:block}. Similar results are obtained for the other ensembles. 
  An alternate estimate of autocorrelation effects can be obtained by calculating the  
  integrated autocorrelation time. We find that the integrated autocorrelation times in the two-point 
 correlation functions included in our analysis are all smaller than 1.4, suggesting that a reasonable block size would be 3 or less.   
We thus choose to account for autocorrelation effects and block the data in all
ensembles by four.  
In a another test, we construct the covariance matrix from the correlation matrix obtained with the unblocked data together with the variances obtained from the blocked data \cite{Bazavov:2017lyh}. Using the same fit setup and parameter choices as before, we find results that are essentially the same as those obtained with our preferred fit method.

\begin{table}
  \caption{Values of the source-sink separation $T$ and $t_\text{min}$ in our preferred
      fits, results for the vector form factor at zero momentum transfer, and one-loop
      finite-volume corrections, $\Delta^Vf_+(0)=f_+^V(0)-f_+^\infty(0)$, on each ensemble---see
    Sec.~\ref{sec:FV} for details of the calculation of $\Delta^Vf_+(0)$. 
    The errors in $f_+(0)$ are statistical only.
    They are generated with a 500-bootstrap distribution.\label{tab:3ptfits}}
  \centering
\begin{tabularx}{0.9\textwidth}{ccc|C{0.2}C{0.08}|C{0.2}|C{0.2}}
\hline\hline
$\approx a({\rm fm})$ & $m_l/m_s^\text{sea}$ & $m_{\pi,P} L$ & $T$ & $t_{min}$ &
$f_+(0)$ & $\Delta^Vf_+(0)$\\
\hline
0.15   & 0.035 & 3.2 & 12,16,17 & 4 & 0.9744(24) & $-0.0007$\\
\hline
0.12   & 0.2   & 4.5 & 15,21,22 & 5 & 0.9874(24) & $0.0002$\\
       & 0.1   & 3.2 & 15,18,21 & 4 & 0.9830(31) & $-0.0003$\\ 
       & 0.1   & 4.3 & 15,18,21 & 4 & 0.9808(22) & $-0.0001$ \\ 
       & 0.1   & 5.4 & 15,18,21 & 4 & 0.9809(17) & $-4\cdot10^{-5}$\\ 
       & 0.035 & 3.9 & 18,21,22 & 6 & 0.9707(18) & $-0.0003$\\ 
\hline
0.09   & 0.2   & 4.5 & 27,32,33 & 3 &  0.9868(18) & $0.0006$\\ 
       & 0.1   & 4.7 & 23,27,32 & 6 &  0.9807(22) & $0.0002$\\ 
       & 0.035 & 3.7 & 23,27,32 & 8 &  0.9709(27) & $-0.0001$\\ 
\hline
0.06   & 0.2   & 4.5 & 34,41,49,50 & 8 &  0.9862(16) & $0.0008$\\ 
       & 0.035 & 3.7 & 31,40,49 & 10 &  0.9697(33) & $0.0005$\\ 
\hline
0.042 & 0.2 & 4.3 & 40,52,53 & 12 & 0.9856(37) & 0.0010\\ 
\hline\hline
\end{tabularx} 
\end{table}

\begin{figure}[tbp]
\centering
\includegraphics[width=0.7\textwidth]{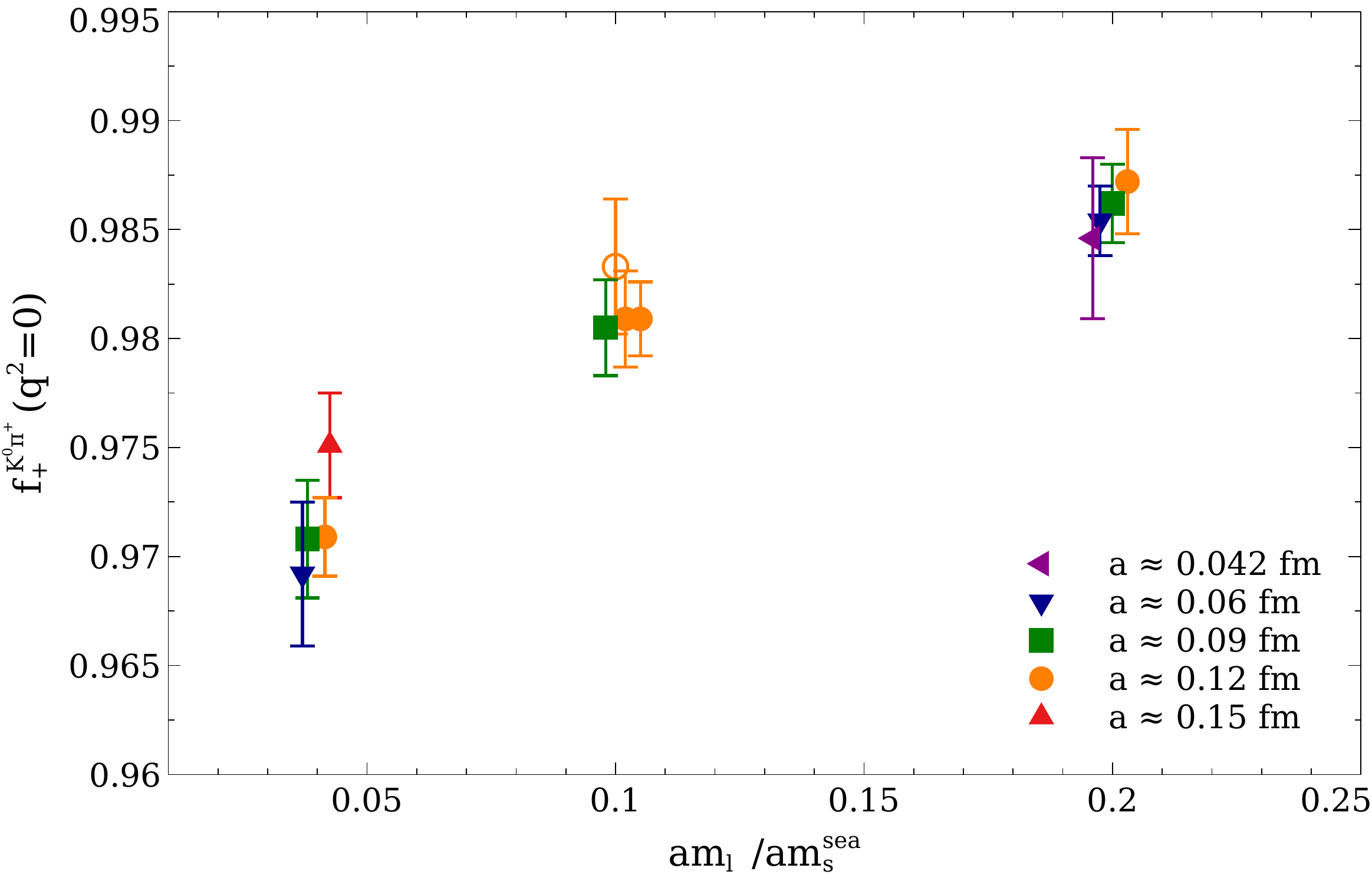}
\caption{
Form factor $f^{K^0\pi^-}(0)$ vs.~light-quark mass. The data points are
the raw results listed in Table~\ref{tab:3ptfits} before applying the
corrections described in Sec.~\ref{sec:corrections}.
    Errors shown are statistical only, obtained from 500 bootstrap resamples.   
    Different symbols and colors denote different lattice spacings. 
    Data at the same light-quark mass but different lattice spacing are offset 
    horizontally. The open orange circle corresponds to the smallest volume ensemble
    with $a\approx0.12~{\rm fm}$ and $m_l/m_s=0.1$. 
\label{fig:latticedata}}
\end{figure}

In Table~\ref{tab:3ptfits} and Fig.~\ref{fig:latticedata}, we show the raw results 
for the vector form factor at zero momentum transfer from the combined fits described 
above. The statistical errors shown in the table and the figure as a function of 
$am_l/am_s^{\rm physical}$ come from 500 bootstrap resamples and range from $0.16\%$ to $0.38\%$. 
The fully correlated covariance matrix is recalculated on each bootstrap resample. 
In the figure, one can see that for a fixed value of the light-quark mass, $m_l$, 
the points with different shapes, which  correspond to different values of the lattice 
spacing, lie on top of each other, with the exception of the data point for the $0.15$~fm ensemble with 
physical quark masses (in the leftmost cluster of points). This is the only ensemble
where we observe statistically significant discretization effects.

%% file: corrections.tex
\section{Form-factor corrections}
\label{sec:corrections}

Before performing the chiral-continuum fit, we correct the form factor results
listed in Table~\ref{tab:3ptfits} and shown in Fig.~\ref{fig:latticedata} for
the leading-order finite-volume effects and the nonequilibrated topological
charge in our finest ensemble. 
These corrections are described in the following two subsections.

\subsection{Finite volume}
\label{sec:FV}

In this work we use NLO ChPT to correct our form factor results for finite-volume effects, whereas in our previous
calculation~\cite{Bazavov:2013maa,Gamiz:2013xxa} we simply estimated the associated systematic error from a comparison 
of the lattice data at two different spatial volumes, with other parameters held fixed. 
The partially twisted boundary conditions used in our calculation introduce several complications
 in the analysis.  In particular, an extra form factor $h_\mu$ is required to parametrize 
the weak-current matrix element in finite volume:
\ba
\langle \pi^-(p')|V_\mu|K^0(p)\rangle = f_+(p_\mu+p'_\mu) + f_- q_\mu + h_\mu.
\ea
The three form factors depend on the choice of twisting angles, as well as 
the value of $q^2$.

We apply the one-loop formulas in Ref.~\cite{Bernard:2017scg}
in the staggered partially-twisted partially-quenched case to all 
ensembles included in our calculation for the choices of twist angles in 
Table~\ref{tab:3ptfits}. Because we are calculating the vector form factor at zero momentum
transfer via  the relation in Eq.~(\ref{eq:WIresult}), and the quantity we obtain at 
finite volume on the lattice is 
$\langle\pi|S|K\rangle(m_s-m_d)/(m_K^2-m_\pi^2)$, 
the FV correction to our results is given by
\ba \label{eq:FVscalar}
\Delta^Vf_+(0)&\equiv& f_+^V(0)-f_+^\infty(0)\nonumber\\
&=&\frac{(m_s-m_d)\Delta^V\langle \pi\vert S
\vert K\rangle}{(m_K^V)^2-(m_\pi^V)^2} - \frac{(m_s-m_d)\langle \pi\vert S
\vert K\rangle^V(\Delta^Vm_K^2-\Delta^Vm_\pi^2)}{\left[(m_K^V)^2-(m_\pi^V)^2\right]^2},
  \ea
where the meson masses in the denominators are the ones from the simulations,
 and quantities that are second order in the finite-volume corrections have been neglected. In \eq{FVscalar}, the FV correction 
 for a given quantity $X$, $\Delta^V X$,  
is defined as $\Delta^VX\equiv X^V-X^\infty$. Notice that, since we extract the meson masses
from correlation functions where all the propagators have zero momentum, the FV
corrections to the meson masses should be calculated from the formulas in
Ref.~\cite{Bernard:2017scg} with zero twisting angles. 

The resulting FV corrections are listed in the last column of Table~\ref{tab:3ptfits}.
We find that they are $\le 0.1\%$ on all ensembles.
Some of the values for $\Delta^Vf_+(0)$ are particularly small due to the cancellation between the two contributions in
Eq.~(\ref{eq:FVscalar}).
We subtract the $\Delta^Vf_+(0)$ from the finite-volume $f_+^{K\pi}(0)$ (listed in the next-to-last column in
Table~\ref{tab:3ptfits}) before performing the chiral-continuum fit discussed in Sec.~\ref{sec:chiral}.

\subsection{Nonequilibrated topological charge}
\label{sec:topo-theory}

The HISQ $N_f=2+1+1$ MILC simulations with smallest lattice spacings have reached a regime where the distribution of the topological
charge $Q$ is not properly sampled~\cite{MILCtopology2010,Bernard:2017npd}, which 
affects the physical observables calculated on those ensembles.
The issue is relevant here for the ensemble with the finest lattice spacing, $a\approx0.042$~fm.
On the other hand, the topological charge is reasonably well equilibrated on the other ensembles, which have $a\gtrsim 0.06$~fm.

In order to correct for this systematic effect, one can use ChPT to study the $Q$-dependence of a given
observable~\cite{Brower:2003yx,Aoki:2007ka,Bernard:2017npd}.
The recent ChPT study in Ref.~\cite{Bernard:2017npd} has already been applied to the calculation of
heavy-light meson decay
constants and masses in Refs.~\cite{Bazavov:2017lyh,Bazavov:2018omf}.
Here, we extend the analysis of Ref.~\cite{Bernard:2017npd} to $f^{K\pi}_+(0)$.

The three-point correlation functions relevant for this study, as well as any meson mass 
calculated in a finite volume $V$ and at fixed $Q$,  
satisfy \cite{Brower:2003yx,Aoki:2007ka}
\ba\label{eq:BQV}
\left.B\right|_{Q,V} = B +\frac{1}{2\chi_TV}B''\left(1-\frac{Q^2}{\chi_TV}\right)
+\order{\left(\frac{1}{\left(\chi_TV\right)^2}\right)},
\ea
where $B$ on the right-hand side is the infinite-volume value of the quantity of interest averaged over $Q$, $B''$ is its second
derivative with respect to the vacuum angle $\theta$, evaluated at $\theta=0$, and
$\chi_T=\lim_{V\to\infty}\langle Q^2\rangle/V$ is the infinite-volume topological susceptibility.
Knowing the dependence on $Q$ or, equivalently, on $\theta$, one can calculate the appropriate correction to $B$ to account for the
difference between the correct $\langle Q^2\rangle$ and the simulation $\langle Q^2\rangle_\text{sample}$.

With Eq.~(\ref{eq:BQV}), 
we follow Ref.~\cite{Bernard:2017npd} to calculate the correction as
\ba\label{eq:fcorr}
\Delta_Q f_+^{K\pi}(0) \equiv f_+^{K\pi}(0)_\text{sample} - f_+^{K\pi}(0)_{{\rm equil}}
=  \frac{1}{2\chi_TV}(f_+^{K\pi}(0))''
\left(1-\frac{\langle Q^2\rangle_\text{sample}}{\chi_TV}\right), 
\ea
where $f_+^{K\pi}(0)_\text{sample}$ is the simulation value.

Although we extract $f_+^{K\pi}(0)$ from the scalar-density matrix element 
in Eq.~(\ref{eq:WIresult}) 
it is simpler to first calculate the $\theta$ dependence of the vector-current matrix element
directly.  In ChPT, the vector current with the relevant flavor is
\ba\label{eq:Vmu-ChPT}
V^\mu=\frac{f^2}{4}\left(\partial^\mu\Sigma\Sigma^\dagger
-\Sigma^\dagger\partial^\mu\Sigma\right)_{13},
\ea
where $\Sigma$ is the SU(3) chiral matrix.  In the presence of $\theta$,
and for the $m_u=m_d=m_l$ and full QCD case (the case relevant for this
work),  the $\order(p^2)$ ChPT Lagrangian is
\ba\label{eq:ChPT-Lag}
{\cal L}_\chi = \frac{f^2}{8}{\rm tr}\left(\partial_\mu\Sigma\partial_\mu\Sigma^\dagger\right)
-\frac{\mu f^2}{4}{\rm tr}\left(\mathcal{M}_A^*\Sigma +  \mathcal{M}_A\Sigma^\dagger\right),
\ea
where $f$ is the chiral-limit value of the meson decay constant, and $\mu$ the 
low energy constant that relates meson and quark masses at leading order (LO)---see Eq.~(\ref{eq:mesonmass}).
Here $\mathcal{M}_A\equiv e^{i\theta/3}\mathcal{M}$, with $\mathcal{M}$ the usual quark mass matrix in the absence of $\theta$.

When $\theta\not=0$, $\Sigma$ gets the vacuum expectation value
\ba
\left\langle \Sigma\right\rangle = \left(\begin{array}{ccc} e^{i\alpha} & 0 & 0\\
  0 & e^{i\alpha} & 0\\ 0 & 0 & e^{-2i\alpha}\\\end{array} \right).
\ea
The parameter $\alpha$ encodes the dependence on $\theta$, with 
$\alpha(\theta\!=\!0)=0$. 
The relation between $\alpha$ and $\theta$ 
is obtained by minimizing the potential energy term in the Lagrangian, which gives the condition
\ba\label{eq:alphacondition}
m_l\sin{\left(\alpha-\frac{\theta}{3}\right)}+m_s \sin{\left(2\alpha+\frac{\theta}{3}\right)}=0.
\ea
For the expansion of the relevant observables, one needs
 $\alpha'$,  the first derivative of $\alpha$ with respect to $\theta$ evaluated
at $\theta=0$. Equation~(\ref{eq:alphacondition})
implies~\cite{Bernard:2017npd}
\ba\label{eq:alphap}
\alpha'=\frac{m_l-m_s}{3(m_l+2m_s)}.
\ea

\begin{figure}[tbp]
\begin{center}
  \includegraphics[width=0.3\textwidth,trim={6cm 21cm 6cm 2.5cm},clip]{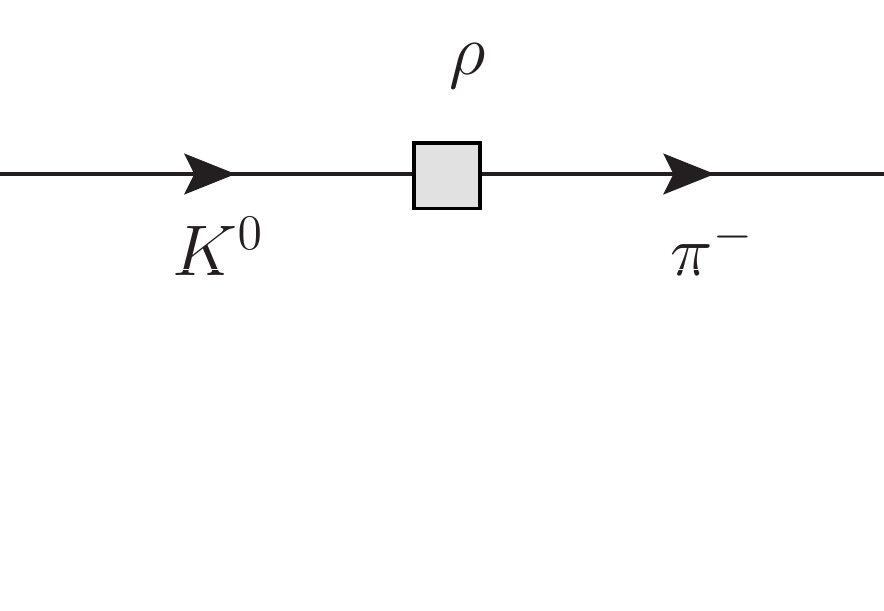}\hspace*{2cm}
  \includegraphics[width=0.3\textwidth,trim={6cm 18cm 6cm 2.5cm},clip]{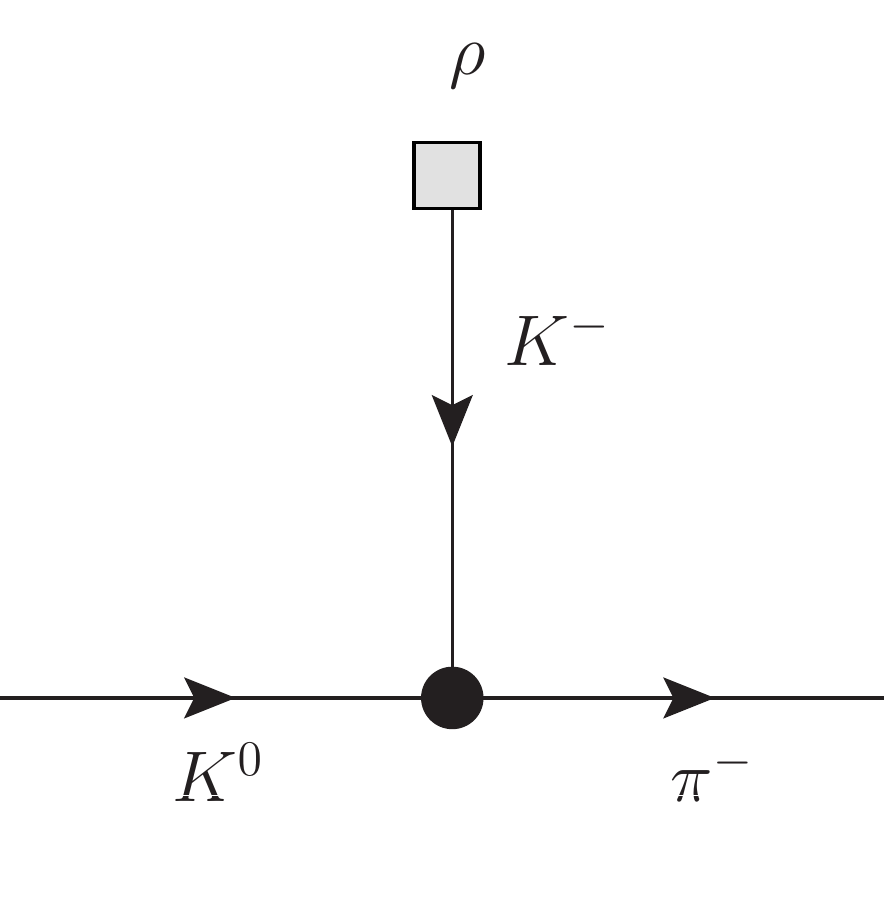}
  \end{center}
    \caption{Diagrams contributing to $\langle \pi\vert \rho \vert K\rangle$ at tree level, where $\rho$ is either the vector
        current~$V^\mu$ or the scalar density~$\tilde S$.
        The squares are weak vertices with the insertion of the current or density and the black dot is a strong vertex. 
    \label{fig:ChPTfortopology}}
\end{figure}

One may expand $\Sigma$ around its vacuum expectation value via
\ba
\Sigma =  \sqrt{\left\langle \Sigma\right\rangle}\; e^{2i\Phi/f}\sqrt{\left\langle \Sigma\right\rangle} ,
\ea
with $\Phi$ the $3\times3$ matrix of meson fields. With this result substituted into Eq.~(\ref{eq:Vmu-ChPT}), 
at tree level there are two possible diagrams, shown in Fig.~\ref{fig:ChPTfortopology},
that  may contribute to
the matrix element $\langle \pi|V^\mu|K\rangle$. The strong three-point vertex in the right-hand diagram is forbidden by
parity when $\theta=0$, but here comes from the mass term in Eq.~(\ref{eq:ChPT-Lag}),
which violates parity symmetry unless one also takes $\theta\to-\theta$ (which is called ``extended
parity'').
The weak vertex in the right-hand diagram generates a factor of $q^\mu$, implying that
that diagram contributes only to the form factor $f_-$.  From the left-hand diagram, one finds
\ba\label{eq:fp-of-alpha}
f_+^{K\pi}(0) = \cos\left(\frac{3}{2}\alpha\right).
\ea
Finally, from Eq.~(\ref{eq:alphap}),  the result needed to adjust the form factor via Eq.~(\ref{eq:fcorr}) is
 \ba\label{eq:fp2deriv}
  f_+^{K\pi}(0)'' = -\frac{1}{4}\frac{(m_l-m_s)^2}{(m_l+2m_s)^2}.
\ea

Because we actually use Eq.~(\ref{eq:WIresult}) to calculate $ f_+^{K\pi}(0)$, it is important
to check that we can reproduce Eq.~(\ref{eq:fp2deriv}) by calculating the
matrix element of the scalar density that appears in the  $\theta\not=0$ Ward identity
at $q^2=0$,
\ba\label{eq:fpvsscalar2}
f_+^{K\pi}(0)=\frac{1}{m_K^2-m_\pi^2}\langle \pi|\tilde{S}|K\rangle_{q^2=0}, 
\ea
with $\tilde S= \frac{1}{2}\bar\psi[\lambda^f,M] \psi$, and $\lambda_f\in SU(N_f)$
the appropriate flavor matrix to select the $\bar s u$ current.
Note that it is no longer convenient to take out a factor of $m_s-m_l$ from $\tilde S$,
as we do for $S$ in Eq.~(\ref{eq:WIresult}), because the quark masses
now carry factors of $\exp(\pm i\theta/3)$. 
In ChPT,
\ba
\tilde S = -\frac{f^2}{4}\mu\left(\Sigma \mathcal{M}_A^*+\Sigma^\dagger\mathcal{M}_A
-\mathcal{M}_A^*\Sigma -\mathcal{M}_A\Sigma^\dagger\right)_{13}.
\ea
Evaluating the diagrams in \figref{ChPTfortopology}, we find
\ba\label{eq:rhofromscalar}
\langle \pi|\tilde S|K\rangle_{q^2=0} && = \mu(m_s-m_l)\cos\left(\frac{\alpha}{2}+\frac{\theta}{3}\right)+\nonumber\\ 
&&\hspace{-2cm}\frac{2}{3}\frac{m_s-m_l}{m_K^2}
\mu^2\sin{\left(\frac{\alpha}{2}+\frac{\theta}{3}\right)} \left[
m_s\sin\left(2\alpha+\frac{\theta}{3}\right)-2m_l\sin\left(\alpha-\frac{\theta}{3}\right)
  \right],
\ea
where the contributions in the first and second lines come from the propagator 
and vertex diagrams in Fig.~\ref{fig:ChPTfortopology}, respectively.

The Ward identity in Eq.~(\ref{eq:fpvsscalar2}) is then satisfied trivially at LO for $\theta=0$.  For $\theta\ne0$,
we may calculate $f_+^{K\pi}(0)''$ from 
Eq.~(\ref{eq:fpvsscalar2}) using \eq{alphap}, the fact that $\alpha''=0$ (which
follows from extended parity), and the second derivatives 
\ba\label{eq:massthetaderiv}
m_\pi'' &=& -m_\pi(0)\frac{m_s^2}{2(m_l+2m_s)^2},\nonumber\\
m_K'' &=& -m_K(0)\frac{m_lm_s}{2(m_l+2m_s)^2} 
\ea
from \cite{Bernard:2017npd}. After some algebra, we find that 
the result agrees with \eq{fp2deriv}, as expected.  We have also checked analytically
that the Ward identity holds for arbitrary $q$ and $\theta$. 

Following the procedure in \rcite{Bernard:2017npd} for the partially quenched case,
we may generalize \eq{fp2deriv} to
\ba\eqn{fp2derivPQ}
f_+^{K\pi}(0)'' = -\frac{1}{4}\frac{m_l^2m_s^2}{(m_l+2m_s)^2}
\frac{(m_x-m_y)^2}{m_x^2 m_y^2},
\ea
where $x$ and $y$ are the active valence quarks (the valence up and strange for 
$f_+^{K^0\pi^-}(0)$), and $m_l$ and $m_s$  are  
the light and strange sea quark masses.  In deriving \eq{fp2derivPQ}, we have
set the spectator  quark mass (the $d$ quark mass for $f_+^{K^0\pi^-}$)
equal to the light sea mass $m_l$; in other words, the spectator quark is
unitary, not partially quenched.   This has allowed us to
avoid analyzing the case of three partially quenched quarks, which was
not treated in \rcite{Bernard:2017npd}.  Since the mass of the spectator quark
does not affect  $f_+^{K^0\pi^-}(0)$ to LO, we believe \eq{fp2derivPQ} will remain
valid even when the spectator quark is partially quenched.  As expected,  
 \eq{fp2derivPQ} reduces to 
Eq.~(\ref{eq:fp2deriv}) when $m_x=m_l$ and $m_y=m_s$.

As discussed at the beginning of this section, the correction is only needed on the finest ensemble included in this analysis,
with $a\approx0.042$ fm. 
We calculate the correction of Eq.~(\ref{eq:fcorr}) using Eq.~(\ref{eq:fp2deriv}), 
the value of the average of the topological charge measured on that ensemble,
$\langle Q^2\rangle_\text{sample}=27.59$~\cite{Bernard:2017npd}, and the correct $\langle Q^2\rangle$ as estimated
by the LO ChPT expression for the topological 
susceptibility~\cite{Leutwyler:1992yt,Billeter:2004wx}
\ba
\chi_T =\frac{f_\pi^2}{4}\left(\frac{1}{2m_{ll,I}^{-2} + m_{ss,I}^{-2}}\right),
\ea
where the singlet meson sea masses are defined in Eq.~(\ref{eq:mesonmass}), below.
The resulting correction $\Delta_Q f_+^{K\pi}(0) =0.00018$ is subtracted from the $f_+^{K\pi}(0)_\text{sample}$ value listed
in the last row of Table~\ref{tab:3ptfits} before performing the chiral-continuum fit.

%% file: chiral.tex
\section{Chiral-continuum interpolation/extrapolation}

\label{sec:chiral}

We follow a methodology very similar to that in our previous analyses~\cite{Bazavov:2013maa,Gamiz:2013xxa} in order to
combine our simulation data into physical results in the continuum limit and with the correct
quark/meson masses.
Here we summarize the main ingredients and then discuss in more detail the new features added in order to accurately account for
finite-volume and isospin-breaking corrections.
Accounting for these effects turns out to be essential, given the improvements in the simulation data.

Our methodology is developed in the framework of chiral perturbation theory (ChPT), which allows us to incorporate effects due to
mass dependence, discretization, finite volume, and isospin breaking in a systematic way.
In particular, in the isospin limit, we can write $f_+^{K\pi}(0)$ as a chiral expansion
\ba
f_+^{K\pi}(0) = 1 + f_2 + f_4 + f_6 + \cdots,
\ea
where the functions $f_i$ are chiral corrections of $\order(p^i)$.
The Ademollo-Gatto (AG) theorem~\cite{AGtheorem} ensures that the vector form factor goes to 1 in the limit $m_s\to m_u$, and that
corrections to this limit are second order.
That means that the functions $f_i$  are proportional to $(m_s-m_u)^2$ or, equivalently, $(m_K^2-m_\pi^2)^2$.
The theorem thus implies that, in the continuum, the $\order(p^2)$ (one-loop) contribution, $f_2$, is completely fixed in terms of
experimental quantities: the decay constant $f_\pi$ and meson masses.

The specific fit function we employ for the extrapolation to the continuum and interpolation to the physical
quark masses is the same as in Ref.~\cite{Bazavov:2013maa}.
It consists of a NLO partially quenched staggered ChPT (PQSChPT)
expression~\cite{Bernard:2013eya} $f_2^\text{PQSChPT}(a)$, plus NNLO continuum ChPT terms~\cite{BT03} $f_4^\text{cont}$, plus
extra analytic terms to parametrize higher-order discretization and chiral effects.
Schematically, it can be written
\ba\label{eq:ChPTtwoloop}
f_+^{K\pi}(0)  =   1   & + &   f_2^\text{PQSChPT}(a) + f_4^\text{cont} + g_{1,a} + 
 r_1^4(m_\pi^2-m_K^2)^2 \left[\tilde C_4 + g_{2,a} + h_{m_\pi}\right] ,
\ea
where the functions $g_{1,a}$ and $g_{2,a}$ account for higher-order discretization effects, and the function $h_{m_\pi}$ includes
analytical terms that parametrize higher-order chiral effects.
We have taken the pure counterterm contribution at two loops out of $f_4^\text{cont}$ and written it separately.
This contribution corresponds to the term proportional to $\tilde C_4$, which is given by the combination of low energy constants
(LECs) $C_{12}+C_{34}-L_5^2$.
The $\order(p^4)$ LEC $L_5$ can be extracted from global fits or from
lattice-QCD calculations of light-light quantities, but
the $\order(p^6)$ LECs $C_{12}$ and $C_{34}$~\cite{Fearing:1994ga,Bijnens:1999sh} are not known.
(Only model-based estimates and imprecise global fit values exist.)
We therefore take $\tilde C_4$ as a constrained fit parameter.
All dimensionful quantities entering in the fit function in Eq.~(\ref{eq:ChPTtwoloop}) are converted into $r_1$ units by using the
values of $r_1/a$ in Table~\ref{tab:inputs}.

\input Tableinputs

Since our simulations are performed in the isospin limit, $m_u=m_d$, $f_2$ and $f_4$ are
evaluated for degenerate up and down quarks.
The explicit NLO PQSChPT function $f_2^\text{PQSChPT}(a)$ can be found in Ref.~\cite{Bernard:2013eya}.
It incorporates the dominant discretization effects coming from the taste-symmetry breaking of staggered fermions. 
The function $f_2^\text{PQSChPT}(a)$ depends on the HISQ taste splittings $\Delta_\Xi$ through the sea meson masses
\ba \label{eq:mesonmass}
m_{ij,\Xi}^2 = \mu (m_i+m_j) + a^2\Delta_\Xi,
\ea 
with $m_i,m_j$ sea quark masses, the slope $\mu$ to be determined by fits of the ChPT expressions to experimentally measured meson
masses, and $\Xi$ labeling the meson taste.
Values of $\Delta_\Xi$ for each ensemble are given in Table~\ref{tab:inputs}.
The function $f_2^\text{PQSChPT}(a)$ also depends on the taste-violating hairpin parameters, $\delta_V'$ and $\delta_A'$, which come
from ChPT disconnected diagrams.
We fix the taste splittings in the fit function to their values in Table~\ref{tab:inputs} since they are precisely enough known that
the corresponding errors do not affect our results significantly.
The values are from Ref.~\cite{HISQensembles}, as well as unpublished updates with better statistics and the inclusion of new
ensembles not previously analyzed.
The uncertainty in the hairpin parameters is, however, quite large.
We therefore treat them as constrained fit parameters with central values and widths equal to those in Table~\ref{tab:priors},
determined from fits to light-light meson quantities \cite{Bazavov:2011fh}.
Their uncertainty is thus propagated to the final fit errors.

\input Tablepriors

For some of the meson masses that appear in $f_2$ there are no experimental measurements or
lattice results, as for example, for $m_{ss}^{\rm valence}$ or for the sea-valence
meson masses involving strange quarks. Because we use values given by NLO ChPT for these
masses, our $f_2$ function has some dependence on the corresponding $\order(p^4)$ LECs
$L_i$. This is the best approximation we have, and we find that different implementations of higher-order corrections result 
in changes to the central values that are significantly smaller than the statistical 
errors.
 
The continuum NNLO ChPT function $f_4^\text{cont}$ also depends on the $\order(p^4)$ LECs.
We take most of them as constrained fit parameters with prior central values equal to the
posteriors obtained in the $\order(p^6)$ global fit BE14 in Ref.~\cite{Bijnens:2014lea}.
We take as an input parameter the combination $2L_6-L_4$ instead of $L_4$ because the fit is more sensitive to that combination and because $L_4$ is fixed in fit BE14.
The prior widths are set to twice the errors in Ref.~\cite{Bijnens:2014lea}. 
The chiral scale, at which the LECs and chiral logarithms in the ChPT expression of
Eq.~(\ref{eq:ChPTtwoloop}) are evaluated, is set equal to the mass of the $\rho$
meson, i.e., $\Lambda_\chi=M_\rho$. 
The $\order(p^4)$ LECs from Ref.~\cite{Bijnens:2014lea}, used as priors here, agree
within errors with (but are more precise than) the only realistic lattice calculations
available at the moment: the $N_f=2+1$ MILC
\cite{Bazavov:2010hj} and the $N_f=2+1+1$ HPQCD \cite{Dowdall:2013rya} calculations.
The prior central values and widths used in our chiral-continuum fit to
Eq.~(\ref{eq:ChPTtwoloop}) are listed in Table~\ref{tab:priors}. 

The $\order(p^4)$ LECs $L_7$ and $L_8$ appear only in the isospin corrections and in
the NLO expressions for some of the meson masses in $f_2$ and $f_4$. 
Their effect on $f_2$ and $f_4$ in the isospin limit is, however, negligibly small,
  and their main impact is via the isospin-breaking corrections, which are added {\em after}
  performing the chiral-continuum fit (see Sec.\ref{sec:isospin}). We choose then to take 
  $L_7$ and $L_8$ as fixed parameters in the chiral-continuum fit and include their
  uncertainties in the total error as described in Sec.~\ref{error:inputs}.

Once the $\order(\alpha_s^2a^2)$ taste-violating discretization errors for staggered fermions
are removed through the explicit dependence on $a$ of  $f_2^\text{PQSChPT}(a)$, the dominant
discretization errors at $\order(p^2)$ in ChPT are $\order(\alpha_s a^2)$ and $\order(a^4)$. 
Since we are forced to use continuum ChPT at $\order(p^4)$, the discretization errors there are
 $\order(\alpha_s a^2)$ and $\order(\alpha_s^2a^2)$. We take these errors into
account through the functions $g_{1,a}$ and $g_{2,a}$ in Eq.~(\ref{eq:ChPTtwoloop}): 
\begin{subequations}
    \label{eq:gas}
\begin{align}
    g_{1,a} = K_1 \sqrt{r_1^2a^2\bar\Delta\left(\frac{a}{r_1}\right)^2} + K_3 \left(\frac{a}{r_1}\right)^4, \\
    g_{2,a} = K_2 \sqrt{r_1^2a^2\bar\Delta\left(\frac{a}{r_1}\right)^2} + K'_2r_1^2 a^2\bar\Delta, 
\end{align}
\end{subequations}
where the $K_i$ are fit parameters, $\bar\Delta =
\frac{1}{16}\left(\Delta_P+4\Delta_A+6\Delta_T+4\Delta_V+\Delta_I\right)$ is the average taste splitting, and 
$r_1^2 a^2 \bar \Delta$ is a proxy for $\alpha_s^2 a^2$. Table~\ref{tab:priors} lists the 
priors employed for the $K_i$s.  
The terms proportional to $K_2$ and $K_2'$ are generic terms parametrizing discretization 
effects of $\order(\alpha_s a^2)$ and $\order(\alpha_s^2a^2)$, respectively, obeying the AG 
theorem.  We include the terms proportional 
to $K_1$ and $K_3$ to account for $\order(\alpha_s a^2)$ and $a^4$ violations of the AG 
theorem at finite lattice spacing arising from symmetry-breaking discretization effects in the 
form factor decomposition, Eq.~(\ref{eq:formfac}), and in the continuum dispersion relation. 
We find that adding an $\order(a^4)$ term instead of the one proportional to $K_2'$ 
yields fit results that are nearly identical.

As in Refs.~\cite{Bazavov:2013maa,Gamiz:2013xxa}, we also add generic analytical terms
corresponding to higher orders in the chiral expansion until the error of the chiral-continuum
fit saturates, i.e., until the central value, the error and the $\chi^2/{\rm dof}$ (and $Q$) value do
  not change appreciably.    
That happens at N${}^4$LO [$\order(p^8)$]---see Sec.~\ref{sec:systematic}.
The function $h_{m_\pi}$ in Eq.~(\ref{eq:ChPTtwoloop}), which collects these effects,
therefore takes the form 
\ba \label{eq:fmpi}
h_{m_\pi} = \tilde C_6\, r_1^2m_\pi^2 + \tilde C_{8}\,r_1^4m_\pi^4.
\ea
The terms proportional to $\tilde C_6$ and $\tilde C_8$ are $\order(p^6)$ and  
$\order(p^8)$, respectively. The $\tilde C_i$  are  constrained fit parameters; the priors 
for them can be found in Table~\ref{tab:priors}. Further discussion of the fit
function, priors used in the Bayesian approach, and tests performed can be found in
Refs.~\cite{Bazavov:2013maa,Gamiz:2013xxa}.

\subsection{Fit results}
\label{sec:fitresults}

\begin{figure}[tbp]
  \centering
\includegraphics[width=0.8\textwidth]{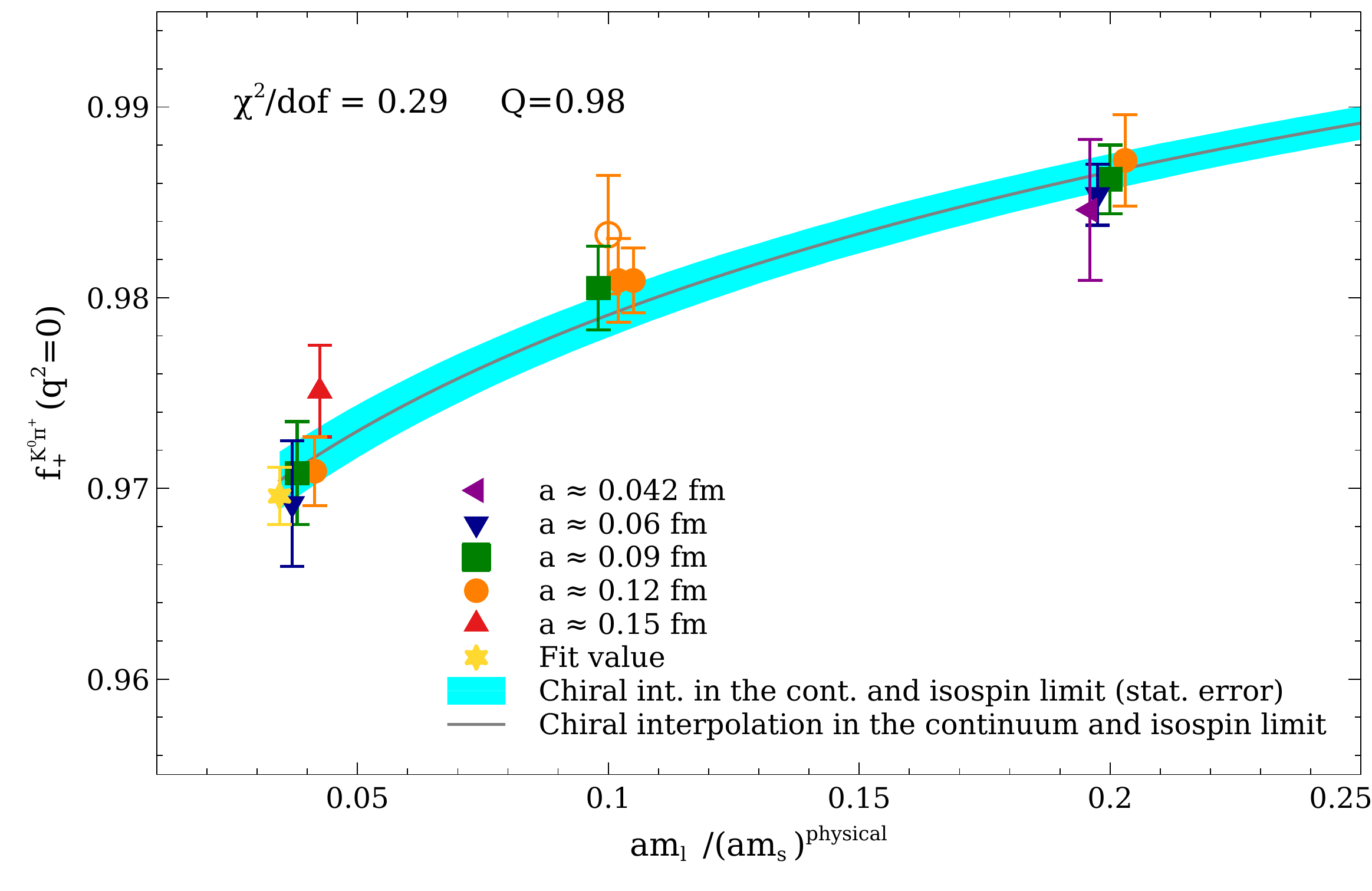}
\caption{
  Form factor $f^{K^0\pi^-}(0)$ vs.~light-quark mass. The data points correspond
    to the results in Table~\ref{tab:3ptfits} and are corrected for the one-loop finite-volume effects also listed in that table.    
    Different symbols and colors denote different lattice spacings.
    The data point at the $a\approx 0.042$~fm ensemble includes the correction given in Sec.~\ref{sec:topo-theory}.
    The error bars on the data points are statistical only, obtained from 500 bootstrap resamples.
    Data points at the same light-quark mass but different lattice spacings are offset horizontally.
    The gray continuous line shows the continuum extrapolation in the isospin limit as a function of the light-quark mass, and the yellow star is the
    continuum result interpolated to the physical light-quark masses.
    The cyan error band, as well as the error bar on the physical point is the statistical chiral-continuum fit error (obtained
    using 500 bootstrap resamples), which includes discretization and higher-order chiral errors, as well as the
    uncertainty from some of the input parameters, as discussed in the text.
    The continuum extrapolation line is obtained by setting the valence and sea light-quark masses equal, setting $m_s$ to its
    physical value, turning off all discretization effects, and considering $m_u=m_d$, i.e., without
      isospin-breaking effects. On the other hand, the yellow star
      is the interpolation to the physical masses and includes strong isospin-breaking effects at NNLO. 
\label{fig:extrapolation}}
\end{figure}

We fit our finite-volume corrected form factor data to the functional form in 
Eq.~(\ref{eq:ChPTtwoloop}) with the functions $g_{i,a}$ and $h_{m_\pi}$ given in
Eqs.~(\ref{eq:gas}) and (\ref{eq:fmpi}), respectively.
All the results listed  in Table~\ref{tab:3ptfits} are included in our central-value fit,
except for the ensemble with $a\approx0.12$~fm and $m_{\pi,P}L=3.2$, 
which we use only to check finite-volume effects. We then extrapolate 
to the continuum limit and interpolate to the pure-QCD meson masses,
i.e., with electromagnetic effects removed, using the parameters determined
from the fit described above together with the continuum isospin-breaking NNLO
ChPT expressions in Ref.~\cite{Bijnens:2007xa}, plus the N${}^3$LO and the N${}^4$LO 
chiral terms in Eq.~(\ref{eq:fmpi}), which do not vanish in the continuum limit.
We take the pure-QCD masses from Ref.~\cite{Bazavov:2017lyh}:\footnote{The $\pi^0$ QCD
  mass is just the experimental one, and the $\pi^+$ QCD mass includes the estimate of the small
  isospin-breaking correction, which comes from Ref.~\cite{Gasser:1984ux}.}
$m_{K^0}^\text{QCD} = 497.567~{\rm MeV}$, 
$m_{K^+}^\text{QCD} = 491.405~{\rm MeV}$,  
$m_{\pi^+}^\text{QCD} = 135.142~{\rm MeV}$,                     
$m_{\pi^0}^\text{QCD} = 134.977~{\rm MeV}$. 
For the $K^0\to\pi^-\ell\nu$ case we find  
\ba\label{fK0pi+}
f^{K^0\pi^-}_+(0) = 0.9696\pm0.0015, 
\ea
where the error is from the fit only, and does not yet include all systematic effects.
The statistical error is estimated by fitting to a set of 500 bootstrap samples for each
ensemble. On each of those fits, we randomly change the central values of all the 
  priors sampling over Gaussian distributions, keeping the same widths as in
  Table~\ref{tab:priors}.  
  The plot in Fig.~\ref{fig:extrapolation} shows, as a function of the light-quark mass, the central
  interpolation curve as well as its error band in the continuum and with the strange-quark mass adjusted
  to its physical value. In order to make the comparison to data clearer, the curve in 
    Fig.~\ref{fig:extrapolation} does not include any strong isospin-breaking effects, i.e., $m_l=m_u=m_d$. 
    The point at the physical masses (yellow star) in Fig.~\ref{fig:extrapolation}, however, is our 
    central result in Eq.~(\ref{fK0pi+}), which includes strong isospin-breaking effects at NNLO.

The result in Eq.~(\ref{fK0pi+}) includes isospin corrections
up to NNLO---see Sec.~\ref{sec:isospin} for more details. 
For $K^0$ decays, isospin corrections enter only at NLO ($f_2$) and beyond,  
and are small, $<0.15\%$. It also includes corrections for the leading-order   
finite-volume effects as described in Sec.~\ref{sec:FV}. 

The second column in Table~\ref{tab:priors} shows the posteriors for the fit parameters  
of the chiral-continuum fit that leads to the result in Eq.~(\ref{fK0pi+}).
We cannot determine the coefficients $K_i$ accurately since 
there is very little $a^2$ dependence in our results. In fact, if we remove the 
$a\approx 0.15$~fm point we could fit our data without including discretization 
effects at all. Our lattice data also provide little constraint on the individual values 
of the $\order(p^4)$ LECs.  As seen in  Table~\ref{tab:priors}, the posterior fit
values of the $L^r_i$ are generally the same as the priors.

%% file: Tableinputs.tex
 \begin{table}[tb]
   \centering
   \caption{Inputs for the parameters taken as fixed in the fit function.  
     The $r_1/a$ values are obtained from a mass-independent scale setting
     \cite{MILCasqtad,HISQensembles}. The absolute scale $r_1$ is from
     Ref.~\cite{Bazavov:2011aa}. The  value of the decay constant $f_\pi$ is 
     taken from Ref.~\cite{Olive:2016xmw}; its error, though shown, is negligible
     in our calculation. Taste splittings $r_1^2a^2\Delta_\Xi$ are taken
      from Ref.~\cite{HISQensembles} and more recent updates;  slopes $a\mu$ come from the
      analysis presented in Ref.~\cite{fermimilcdecay2014}, although they were not published
      there. We do not consider errors either on the taste splittings or on the slopes 
      because they also have a negligible effect on the final results. Notice that taste
      splittings for the $a\approx 0.042$~fm ensemble are not measured but obtained from the
      $0.06$~fm results, by applying the expected scaling factor $\alpha_s^2 a^2$.
      The LECs $L_7$ and $L_8$, both central values and errors, are taken from fit BE14 in Ref.~\cite{Bijnens:2014lea}.} 
\label{tab:inputs}
\begin{center}  
  \begin{tabular}{ccccccc}
    \hline\hline
$\approx a$ (fm) & 0.15 & 0.12 & 0.09 & 0.06 & 0.042 & continuum \\
\hline
$r_1$   & & & & & &$0.3117\pm0.0022$~fm \\
$f_\pi$ & & & & & &$130.50\pm0.13$~MeV  \\
$\Lambda_{\chi} r_1=M_\rho r_1$ & & & & & & 1.2163 \\
    \hline
$a\mu$ & 2.0565 & 1.6994 & 1.2820 & 0.8873 & 0.6986 & \\
\hline
$r_1/a$ & 2.090(6) & 2.608(4) & 3.588(7) &  5.442(10) &
7.143(24) & \\
\hline
$r_1^2a^2\Delta_P$ & 0 & 0 & 0 & 0 & 0 & \\
$r_1^2a^2\Delta_V$ & 0.301197 &0.167563 &0.052723 &0.009542 &0.004794 & \\
$r_1^2a^2\Delta_T$ & 0.204127 &0.103326 &0.034894 &0.006974 &0.003504 & \\
$r_1^2a^2\Delta_A$ & 0.106046 &0.053983 &0.018187 &0.003588 &0.001803 & \\
$r_1^2a^2\Delta_I$ & 0.399862 &0.209269 &0.066393 &0.012493 &0.006276 & \\
\hline
$L_7^r(\Lambda_\chi)\times 10^3$ & & & & & & $-0.34\pm0.09$ \\
$L_8^r(\Lambda_\chi)\times 10^3$ & & & & & & $\hphantom{-}0.47\pm0.10$ \\
\hline\hline
\end{tabular}
\end{center}
 \end{table}

%% file: Tablepriors.tex
\begin{table}[tb]
    \newcommand{\m}{\hphantom{-}}
\caption{
  Priors for the fit parameters entering in Eq.~(\ref{eq:ChPTtwoloop}),
  as well as the posterior values obtained for those
  parameters in our preferred fit. The dimensionless $\chi$PT
  parameter $s$ is given by the quantity $1/(8\pi^2(r_1f_\pi)^2)\approx0.3$.
  The priors listed for the hairpin parameters are for the
  $a\approx0.12~{\rm fm}$ ensembles, and those for the other lattice spacings
  are obtained by rescaling these numbers, assuming that the hairpin
  parameters scale like the average of the $\Delta_\Xi$. These values are
  obtained from fits to light-light quantities using two-loop
  PQChPT~\cite{Bazavov:2011fh}. The uncertainty includes statistical and
  systematic errors. The prior central values for the NLO LECs are from fit BE14 in
  Ref.~\cite{Bijnens:2014lea} with $\Lambda_\chi=0.77~{\rm GeV}$, while the
  prior widths are twice the errors in Ref.~\cite{Bijnens:2014lea}. We fix the
  LECs $L_7$ and $L_8$ and give their values in Table~\ref{tab:inputs}, as
  explained in the text. 
  The entries ``$-0.000$'' denote small negative numbers that round to zero.}
   \label{tab:priors}
    \centering
\centering
\begin{tabular}{ccc}
\hline\hline
Fit parameters & Gaussian priors & ChPT fit \\
& (central value $\pm$ width) & posteriors \\
\hline
$r_1^2a^2\delta_V'$ & $\m0.050\pm0.024$ & $\m0.050\pm0.024$\\
$r_1^2a^2\delta_A'$ & $-0.0946\pm0.0094$ & $-0.0958\pm0.0093$\\
\hline
$K_1$ & $0\pm 0.01$ & $\m0.001\pm 0.010$\\
$K_2$ & $0\pm0.03$ & $\m0.001\pm 0.030$\\
$K_2'$ & $0\pm0.81$ & $\m0.083\pm 0.063$\\
$K_3$ & $0\pm0.015$ & $-0.000\pm 0.015$\\
$\tilde C_4$ & $0\pm s^2$ & $-0.052\pm0.006$\\
$\tilde C_6$ & $0\pm s^3$ & $\m0.006\pm0.022$\\
$\tilde C_8$ & $0\pm s^4$ & $-0.000\pm0.008$\\
\hline
$L_1^r(\Lambda_\chi)\times 10^3$ &  $\m0.53\pm0.12$  & $\m0.55\pm 0.12$\\
$L_2^r(\Lambda_\chi)\times 10^3$ &  $\m0.81\pm0.08$  & $\m0.81\pm 0.08$\\ 
$L_3^r(\Lambda_\chi)\times 10^3$ &  $-3.07\pm0.40$ & $-3.03\pm 0.40$\\ 
$[2L_6^r-L_4^r(\Lambda_\chi)]\times 10^3$ &  $-0.02\pm0.10$ & $-0.01\pm 0.11$ \\
$L_5^r(\Lambda_\chi)\times 10^3$ &  $\m1.01\pm0.12$  & $\m1.00\pm 0.12$\\
$L_6^r(\Lambda_\chi)\times 10^3$ &  $\m0.14\pm0.10$  & $\m0.13\pm 0.09$\\
\hline\hline
\end{tabular}
\end{table}

%% file: systematic.tex
\section{Systematic error analysis}

\label{sec:systematic}

The error in Eq.~(\ref{fK0pi+}) includes statistical, chiral-extrapolation, and discretization errors, as well as
the uncertainties associated with the inputs that are treated as constrained fit parameters: $\order(p^4)$ LECs (except $L_{7,8}$)
and taste-violating hairpin parameters. 
The uncertainties of the data and constrained input parameters are propagated through the fit via 500 bootstrap resamples.

In this section, we further study these sources of 
uncertainty, perform tests of the stability of our preferred fit strategies, and 
estimate the other sources of systematic error entering in our calculation of 
$f_+^{K\pi}(0)$: uncertainty in the inputs, scale error, partial-quenching effects,
higher-order finite-volume effects, isospin-breaking corrections, and the effects of nonequilibrated 
topological charge.

\subsection{Fit function, discretization error and chiral interpolation}
\label{sec:statistical}

\begin{figure}[tbp]
\includegraphics[width=1.\textwidth]{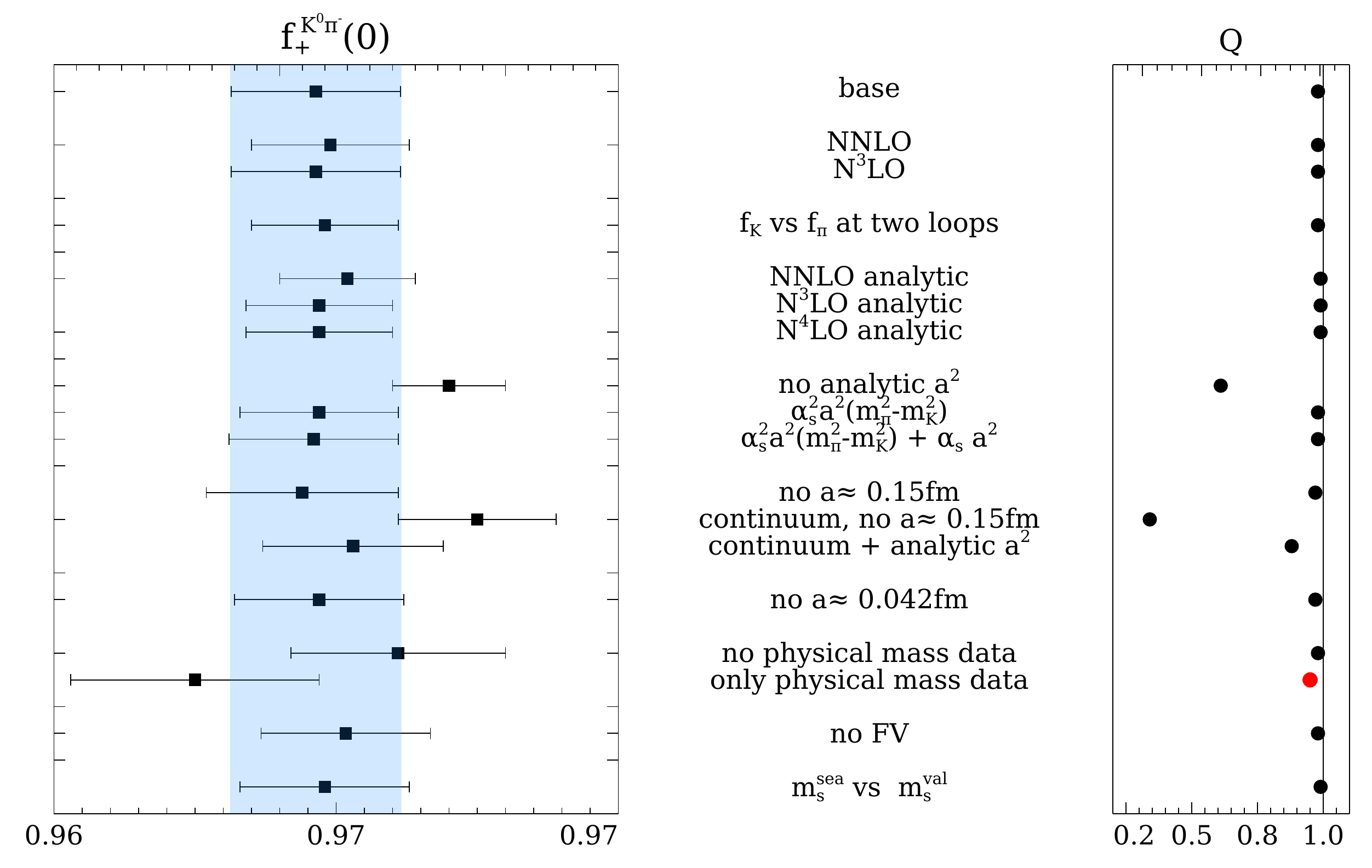}
\caption{
  Stability of the continuum extrapolation and chiral interpolation with respect to the choice of fit function. 
The blue band corresponds to our preferred fit function (labeled ``base'').
Notice that the analytical parametrization is applied only at
NNLO and beyond. The PQSChPT expression is used at NLO, including isospin
corrections. The $Q$ value for each fit is shown in the right-hand-side plot. 
The red point on that plot corresponds to a fit with $\chi^2/dof<0.05$.  
See the text for the explanation of the different tests performed.}
\label{fig:fitchecks} 
\end{figure}

Because we have data at the physical light-quark masses, the chiral fit is an interpolation, and is largely independent of the
precise form of the fit function and the values of the ChPT parameters.
We have performed a number of tests to check this stability under variations in the fit function and to estimate the effect of
higher-order terms in the chiral and Symanzik expansions.
Figure~\ref{fig:fitchecks} shows the tests performed and discussed in this subsection, together with the results from additional
fits discussed in the following subsections, which we use to estimate several systematic uncertainties.

First of all, in order to be sure that effects due to higher-order terms in the chiral
expansion are properly included in the Bayesian analysis that leads
to the fit error shown in Eq.~(\ref{fK0pi+}), we need to check that 
this error stabilizes as we add higher-order chiral terms. 
The point labeled NNLO in Fig.~\ref{fig:fitchecks} includes only terms up to NNLO, i.e.,
without the $h_{m_\pi}$ function in Eq.~(\ref{eq:ChPTtwoloop}). 
Only minimal changes in the central value and errors are produced by the addition of the N${}^3$LO term
$\tilde{C}_6(m_K^2-m_\pi^2)^2m_\pi^2$ from Eqs.~(\ref{eq:fmpi}) and (\ref{eq:ChPTtwoloop}).
The difference between this and the base fit is the N${}^4$LO term $\tilde C_8(m_K^2-m_\pi^2)^2m_\pi^4$,
and we see that it has a negligibly small effect. 

Different values of the decay constant used in the two-loop (NNLO) term $f_4^\text{cont}$ are equivalent up to omitted higher-order
terms, and therefore should have a negligible effect on our analysis.
In contrast, the decay constant at one loop has to be set equal to the physical pion decay constant, $f_\pi$, in order to be 
consistent with the particular expression chosen for $f_4^\text{cont}$.
In the base fit, we use $f_\pi$ as the chiral expansion parameter in $f_4^\text{cont}$.
We check that other possible choices, such as $f_K=(155.6\pm0.4)$~MeV~\cite{Rosner:2015wva} (labeled ``$f_K$ vs.\ $f_\pi$ at two
loops'' in Fig.~\ref{fig:fitchecks}) and an estimate of the decay constant in the chiral limit, $f_0=(113.5\pm8.5)$~MeV~\cite{Bazavov:2010hj}, shift the central value in Eq.~(\ref{fK0pi+}) by less than 0.06\%, well under the
statistical error.

As another test of the ChPT fit and errors, we replace the continuum two-loop ChPT expression in Eq.~(\ref{eq:ChPTtwoloop}),
$f_4^\text{cont}$, by an analytic function, and consecutively add N${}^3$LO and N${}^4$LO analytic
terms---see results labeled ``NNLO analyt.,'' ``N$^3$LO analyt.,'' and ``N$^4$LO analyt.,'' respectively, in
Fig.~\ref{fig:fitchecks}.
All the results agree very well with our base fit within statistics, being nearly identical once the N${}^3$LO analytic term is
included.

The three results labeled ``no analyt.~$a^2$'' (which corresponds to a fit without including $g_{1,a}$ and $g_{2,a}$ in the fit
function), ``$\alpha_s^2a^2(m^2\pi-m^2_K)$'', and ``$\alpha_s^2a^2(m^2_\pi-m^2_K)+\alpha_s a^2$'' in \figref{fitchecks} represent a
check that the discretization errors are properly included in the fit error of Eq.~(\ref{fK0pi+}).
Once we include the term of order $\alpha_s^2a^2(m_K^2-m_\pi^2)^2$ (proportional to $K'_2$) in Eq.~(\ref{eq:gas}),
which is required to get a fit of similar quality to our base fit 
  ---see Fig.~\ref{fig:fitchecks}, the central value and errors barely change with the addition of
  $\alpha_s a^2$ corrections (the result labeled ``$\alpha_s^2a^2(m^2_\pi-m^2_K)+\alpha_s a^2$'').
Adding the two remaining discretization terms in $g_{1,a}$ and $g_{2,a}$, which returns us to the base fit, makes no noticeable
difference.
The rapid stabilization of the fit reflects the tiny lattice-spacing dependence of our data.

Of all the data in \figref{latticedata}, only the point at $a\approx0.15$~fm shows what appear to be significant
discretization effects.
Dropping that data point has the effect of increasing the errors (see result labeled ``no $a\approx0.15$ fm''), since the other
ensembles provide very little constraint of the analytical $a^2$ fit parameters.
In fact, after dropping that point, we can fit our remaining data with a continuum fit function, although we see from
\figref{fitchecks} (result labeled ``continuum, no $a\approx 0.15$~fm'') that the result is larger than our central result by 
about two standard deviations, measured in terms of the fit errors, and the quality of the fit significantly drops.
Adding analytical discretization corrections via the functions $g_{1,a}$ and $g_{2,a}$ to the continuum fit function allows us to
fit all our data, giving a result that is consistent with the base fit and with a similar $Q$ value 
(see result labeled ``continuum + analyt. $a^2$''), although with a larger error.
    
In contrast to the noticeable effect of the coarsest ensemble on the total error, the effect of our finest lattice spacing,
$a\approx0.042$~fm, on the central value and the error is very small since statistics in this ensemble is limited and, in addition,
it has $m_l=0.2m_s$, so it is relatively far from the physical point.

As shown in Fig.~\ref{fig:fitchecks}, both the ensembles with physical quark masses 
and those with unphysical masses are important in fixing the central value and
reducing the fit error. The larger error of the fit including only physical-quark-mass 
ensembles  reflects the weaker constraints on the higher-order discretization terms
and the  lack of constraints on the
higher-order chiral terms, which can have an effect on the results from
nominally ``physical''  ensembles due to mistunings of the strange and light-quark masses. 
On the other hand, the larger error of the fit including only the unphysical-quark-mass 
ensembles  reflects primarily the error of the chiral extrapolation.

Finally, we test the robustness of our Bayesian error estimation strategy 
  similarly to our previous work~\cite{Bazavov:2013maa,Gamiz:2013xxa}, by obtaining separate
  estimates of each source of error from central value variations observed with simpler fits with and
  without the corresponding higher-order terms---see 
  Ref.~\cite{Bazavov:2013maa} for details. Taking the total error as their quadrature sum,
  we find that this procedure yields smaller uncertainties than those in Eq.~(\ref{fK0pi+}).

For the reasons discussed above, the statistical fit error shown in Eq.~(\ref{fK0pi+}), 
which is obtained with our base fit 
using Eq.~(\ref{eq:ChPTtwoloop}), together with the higher-order
chiral and discretization terms in Eqs.~(\ref{eq:gas}) and~(\ref{eq:fmpi}), properly includes the errors from higher-order
discretization effects and chiral corrections in addition to the statistical errors.
The inclusion of the unphysical light-quark-mass data in our ChPT description gives us a handle on these higher-order
effects and allows us to robustly correct for mass mistunings and estimate the error associated with the truncation of the
corresponding series.

\subsection{Inputs for the fixed parameters in the chiral function}
\label{error:inputs}

The values and errors of the fixed inputs we use in our chiral-continuum  
fit are listed in Table~\ref{tab:inputs}. The HISQ taste 
splittings are known precisely enough that their errors have no impact on
the final uncertainty. 
Similarly, when we change the pion decay constant within its error and repeat
the fit, results for the form factor are unchanged at the precision we quote.
The uncertainty is small because the dependence on~$f_\pi$ enters through the
coefficients and parameters in the ChPT fit function, which, as discussed above,
already have little effect on the results.
Finally, by varying $\Lambda_\chi$ in the range $M_\rho\pm0.5~{\rm GeV}$, we have
checked that our results are independent of the chiral scale, as they should be.
We therefore do not need to add an uncertainty due to the errors in the inputs or 
the choice of chiral scale to the statistical fit error. 

However, the LECs $L_7$ and $L_8$ (which we treat as fixed input parameters, unlike the other LECs),
do have an impact on the form factor error, mainly through their effect on the
isospin corrections. 
We estimate this uncertainty by varying their central values by their respective standard deviations,
repeating the fit, and recalculating the form factor (including isospin corrections).
We take the shift that this variation produces as the uncertainty associated with these LECs,
and add it in quadrature to the fit error, as shown in Table~\ref{tab:errorbudget}. 
The above procedure does not underestimate the error due to these LECs, since if we treat $L_7$
and $L_8$ as constrained fit parameters instead, the same as the other $\order(p^4)$ LECs, we
obtain a slightly smaller total error.

\subsection{Lattice scale}

\label{sec:scale}

We rewrite all the dimensionful quantities entering in the two-loop ChPT fit
function in $r_1$ units, where the $r_1$ scale is obtained from the static-quark 
potential~\cite{r11,r12}. The lattice parameters are converted to $r_1$ units by 
multiplying by the values of the relative scales $r_1/a$ in Table~\ref{tab:inputs}, 
while the physical parameters are converted by using 
$r_1=0.3117(22)~{\rm fm}$~\cite{Bazavov:2011aa}. 

The form factor $f_+^{K\pi}(0)$ is a dimensionless quantity, and thus the effect of the 
error in the lattice scale is small. When we change the scale $r_1$
by its error, the central value only shifts by $\pm 0.0008$. 
We include this variation as a systematic error in Table~\ref{tab:errorbudget}. 
The errors in the relative scales $r_1/a$, on the other hand, have no significant
impact on our results.

\subsection{Partial-quenching effects in $m_s$ at NNLO}

The valence and sea strange-quark masses differ on some of the ensembles
as explained in Sec.~\ref{sec:simulations}, leading to partial-quenching
effects---see Table~\ref{tab:ensembles}. These effects can be exactly treated
at NLO within the PQSChPT framework, but at NNLO only the full-QCD ChPT
expressions are available. We then have the choice of using either the sea or
the valence $m_s$ at NNLO and beyond. In practice, this ambiguity only affects
$f_4^\text{cont}$ in Eq.~(\ref{eq:ChPTtwoloop}) since the factor
$(m_\pi^2-m_K^2)^2$ in that equation comes from the valence sector. 

The result in Eq.~(\ref{fK0pi+}) is  obtained using the
valence strange-quark masses at NNLO. If we use the sea strange-quark masses
at NNLO instead, $f_+^{K^0\pi^-}(0)$ shifts by $0.013\%$,  
which we include on the line labeled ``$m_s^{\rm val}\ne m_s^{\rm sea}$''
in the error budget. This systematic effect is small because the sea strange-quark 
masses are generally well tuned on the HISQ $N_f=2+1+1$ MILC ensembles, and 
$m_s^{\rm val}= m_s^{\rm sea}$ on the most relevant ensembles in the chiral-continuum 
interpolation/extrapolation, the ensembles with physical quark masses and 
$a\approx 0.09,0.06$~fm.

\subsection{Higher-order finite-volume corrections}

\label{sec:FVsys}

In our previous calculation~\cite{Bazavov:2013maa,Gamiz:2013xxa}, the uncertainty due to finite-volume effects was one of the two
dominant sources of error.
(The other was the fit error.) The finite-volume error was estimated to be of the same order as the statistical error from a
comparison of the lattice data from two different volumes, with other parameters held fixed.
Then, although very small, $0.2\%$, this error turned out to be a limiting factor for precision.
In this work we have increased the statistics on the ensembles analyzed in Refs.~\cite{Bazavov:2013maa,Gamiz:2013xxa} to check finite
size effects.
We have also sharpened this direct comparison by generating data on a third, smaller, volume.
The three ensembles are those with $a\approx0.12$~fm and $m_l/m_s^{{\rm sea}}=0.1$ in Table~\ref{tab:ensembles} and
Fig.~\ref{fig:ensembles}.
Table~\ref{tab:3ptfits} gives the values for $f_+^{K\pi}(0)$ on these three volumes.
The results on the two largest volumes are essentially the same, while that on the smallest volume differs from the others by less
than the statistical error.
From this comparison alone we could conclude that finite-volume effects are smaller than 0.17\%, the smallest statistical error on
the three ensembles.

To reduce the error further we use NLO staggered partially-twisted partially-quenched ChPT~\cite{Bernard:2017scg} to correct the
form factor prior to the chiral-continuum fit, as described in Sec.~\ref{sec:FV}.
The resulting finite-volume corrections are $\le 0.1\%$ on all ensembles.
If we did not correct our data for finite-volume effects at one loop, the result for $f_+^{K^0\pi^-}(0)$ would shift by 0.00051. 
Although we expect NNLO finite-volume corrections to be suppressed by a typical
  one-loop suppression factor, we conservatively take this shift as the estimate for the
  higher-order finite-volume effects. This gives a 0.053\% error that we include in
  the error budget in Table~\ref{tab:errorbudget}. 

\subsection{Isospin-breaking corrections}

\label{sec:isospin}

Isospin-breaking corrections accounting for the difference between the up- and
down-quark masses can be calculated in the ChPT framework and thus written as a
chiral expansion starting at NLO for neutral kaons
\ba\label{eq:isospincorrections}
\Delta_{\rm isospin}f^{K^0\pi^-}_{+}(0) \equiv f^{K^0\pi^-}_{+}(0)
  -f^{K\pi}_{+,{\rm isospin\,limit}}(0)  = & \sqrt{3}\left(\zeta_{S,K^0\pi^-}^{(4)} + 
\zeta_{S,K^0\pi^-}^{(6)}\dots\right), 
\ea
where the  parameters $\zeta_{S,K^0\pi^-}^{(i)}$ are $\order((m_u-m_d)p^{i})$ isospin
corrections. In our result for $f^{K^0\pi^-}_+(0)$ in Eq.~(\ref{fK0pi+}), we include 
both NLO [$\order((m_u-m_d)p^{4})$] and NNLO [$\order((m_u-m_d)p^{6})$] corrections
calculated in Refs.~\cite{Gasser:1984ux} and \cite{Bijnens:2007xa}, respectively. 
These corrections depend on the lowest-order $\pi^0-\eta$ mixing angle
$\varepsilon^{(2)}$, or, alternatively, the quantity 
$R\equiv(m_s-\hat m)/(m_d-m_u)$ with $\hat m\equiv (m_u+m_d)/2$. In order to arrive at the
number in Eq.~(\ref{fK0pi+}), we use the expressions in Ref.~\cite{Bijnens:2007xa}, 
the QCD meson masses quoted in Sec.~\ref{sec:fitresults}, and the values of the LECs
obtained from our fits and shown in Table~\ref{tab:priors} (for $L_7^r$ and $L_8^r$ we
take the input values in Table~\ref{tab:inputs}). The only combination of $\order(p^6)$
LECs that enters at this order in the isospin-breaking terms for $K^0\to\pi^- \ell\nu$
decays is $C_{12}+C_{36}$. This combination, which we obtain from our
fitting procedure, is the same one that appears in the isospin limit.

We use a power-counting estimate for the error due to isospin corrections not included
in our result, N${}^3$LO and higher, by taking the calculated
NNLO correction and multiplying it by a typical chiral-loop suppression factor.
For quantities involving a strange quark, we may estimate this factor to be
$m^2_K/(8\pi^2f_\pi^2)\approx0.18$. The size of the ratio of the isospin limit
NNLO and NLO contributions to $f_+^{K^0\pi^-}(0)$ that we obtain in this work is a bit
larger, $\approx0.26$. We conservatively multiply the calculated 
NNLO isospin-breaking correction, $-0.00057$, by the larger number,
  which yields a  0.015\% uncertainty.  

Another source of error is the parametric uncertainty in the isospin-breaking quantity~$R$ used to obtain the corrections in
Eq.~(\ref{eq:isospincorrections}). We use the value
\ba
  R = 35.59 (21)_\text{stat}({}^{+88}_{-96})_\text{syst}[35]_\text{EM-scheme}.
\ea
The analysis that yields to this result is the same as in Ref.~\cite{Bazavov:2017lyh}, except that we have included more
configurations at the ensembles with $a\approx 0.06~{\rm fm}$ and $a\approx 0.042~{\rm fm}$, and included the $a\approx 0.15~{\rm fm}$ data in the central fit.
The electromagnetic errors are estimated as in Ref.~\cite{Basak:2018yzz}.

We estimate the error on the form factor coming from the uncertainty on $R$ by varying this
quantity within its error and redoing the fit.
As expected, the impact on the form factor for the neutral mode is nearly negligible, $0.002\%$.
Nevertheless, we include it in our error budget.

\begin{table}[tbp]
  \caption{Error budget for $f_+^{K^0\pi^-}(0)$ in percent.
       \label{tab:errorbudget}}
  \centering
    \begin{tabular}{lc}
    \hline\hline
    Source of uncertainty & Error $f_{+}^{K^0\pi^-}(0)$ (\%)\\ 
    \hline
    Statistical + discretization + chiral interpolation & $0.154$\\
    $L_{7,8}^r$ & $0.079$ \\
    Scale $r_1$ & $0.080$ \\  
    $m_s^{\rm val}\ne m_s^{\rm sea}$ & $0.013$ \\
    Higher-order finite-volume corrections & $0.053$ \\ 
    Higher-order isospin corrections & $0.015$ \\ 
    Isospin-breaking parameter $R$ & $0.002$ \\
    \hline
    Total Error & $0.199$ \\
    \hline\hline
    \end{tabular}
\end{table}

\subsection{Nonequilibrated topological charge}
\label{sec:topo-systematic}

As described in Sec.~\ref{sec:topo-theory}, a correction due to improper sampling of the topological charge is needed only on the
$a\approx 0.042$~fm ensemble with $m_l=0.2m_s$, where we obtain $\Delta_Q f_+^{K\pi}(0) =0.00018$.
Not surprisingly, given that (i) this ensemble has little influence on the chiral-continuum interpolation/extrapolation (see
Fig.~\ref{fig:fitchecks} for the effect of removing the ensemble completely), and (ii) the correction is much smaller than the
statistical error on the ensemble (see Table~\ref{tab:3ptfits}), the effect of the correction on the physical value of
$f_+^{K\pi}(0)$ is negligible.
We therefore do not add an uncertainty due to this effect to our error budget.

%% file: results.tex
\section{Results}
\label{sec:results}

Our final result for the vector form factor is
\ba\label{eq:fplusfinal}
f_+^{K^0\pi^-}(0) & = & 0.9696(15)_\text{stat}(12)_\text{syst}  = 0.9696(19),
\ea
where the first error in the middle expression is the combined statistical, discretization and chiral interpolation error discussed
in Sec.~\ref{sec:fitresults}, and the second the sum in quadrature of all the systematic errors discussed in
Sec.~\ref{sec:systematic}.
Table~\ref{tab:errorbudget} summarizes all sources of error in our calculation.
The total uncertainty is the smallest achieved to date. 

\begin{figure}[tbp]
\begin{center}
\includegraphics[width=0.75\textwidth]{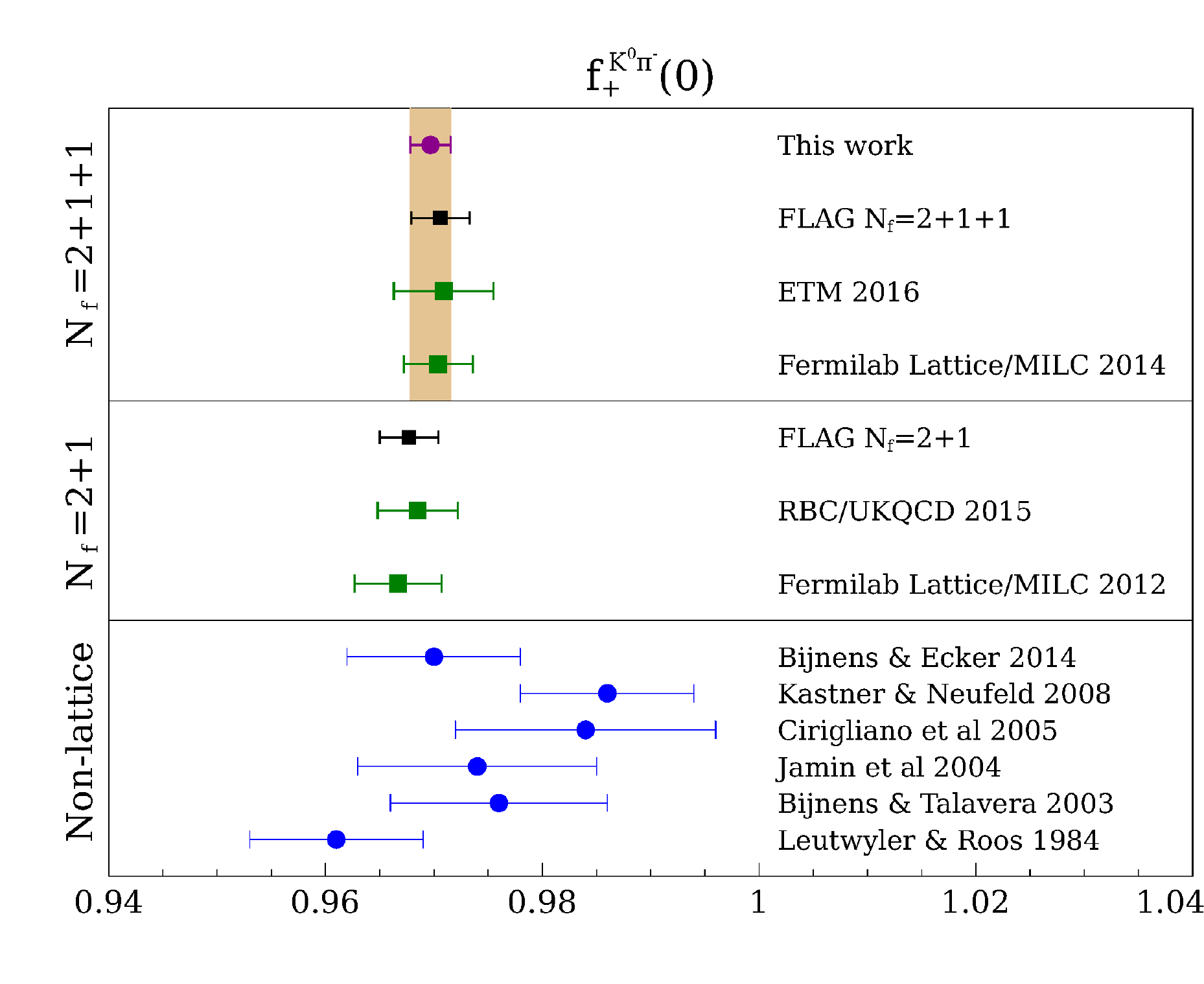}
\end{center}
\caption{Comparison of $f_+^{K^0\pi^-}(0)$ from this analysis with previous lattice results entering 
  in the FLAG averages~\cite{Aoki:2016frl} together with those averages for $N_f=2+1+1$
  and $N_f=2+1$, as well as nonlattice determinations based on ChPT. The beige band
  corresponds to our result. The references and numerical results for all
  determinations are given in Table~\ref{tab:comparison}.
  \label{fig:Resultsf+}}
\end{figure}

We compare our result for $f_+^{K^0\pi^-}(0)$ with the results from the most recent lattice
calculations and phenomenological approaches in Table~\ref{tab:comparison},
and with the results entering the FLAG average and those from 
phenomenological approaches in Fig.~\ref{fig:Resultsf+}.
Our value for $f_+^{K^0\pi^-}(0)$ agrees within errors with previous $N_f=2+1$ and $N_f=2+1+1$ lattice
calculations. In particular, the value is close to the other $N_f=2+1+1$ results, 
but with significantly smaller errors. It also agrees with the most recent
phenomenological determinations \cite{Bijnens:2014lea,Ecker:2015uoa}, which are
based on two-loop ChPT with LECs determined by NNLO global fits. 
The lattice results in Table~\ref{tab:comparison} and in Fig.~\ref{fig:Resultsf+} do not
include isospin corrections, with the exception of the Fermilab
Lattice/MILC result in 
Ref.~\cite{Bazavov:2013maa} (only NLO corrections) and our result here (up to NNLO
corrections).

\begin{table}[tbp]
    \caption{Form factor $f_+^{K^0\pi^-}(0)$ as extracted from the most recent lattice calculations
    (first half of the table), from phenomenological approaches using of two-loop ChPT, and
    from the 1984 calculation by Leutwyler and Roos, which uses one-loop ChPT and a
    quark model for higher-order terms. For those calculations based on two-loop ChPT,
    we also indicate the method used in the estimate of the $\order(p^6)$ LECs. 
\label{tab:comparison}}
  \begin{tabular}{lll}
\hline\hline
Group & $~~f_+^{K^0\pi^-}(0)$ & \qquad Method \\
\hline
This work & $0.9696(15)(12)$ & staggered fermions ($N_f=2+1+1$) \\
ETM~\cite{Carrasco:2016kpy}     & $0.9709(45)(9)$ & twisted-mass fermions ($N_f=2+1+1$)\\
Fermilab Lattice/MILC~\cite{Bazavov:2013maa} & $0.9704(24)(22)$ & staggered fermions ($N_f=2+1+1$) \\
JLQCD~\cite{Aoki:2017spo} & $0.9636(36)(^{+57}_{-35})$ & 
overlap fermions ($N_f=2+1$) \\
RBC/UKQCD~\cite{Boyle:2015hfa} & $0.9685(34)(14)$ & domain-wall fermions ($N_f=2+1$) \\
\hline
Bijnens \& Ecker~\cite{Bijnens:2014lea,Ecker:2015uoa} & 0.970(8) &  ChPT + NNLO global fit \\
Kastner \& Neufeld~\cite{Kastner:2008ch} & $0.986(8)$ &  ChPT + large $N_c$ + dispersive \\
Cirigliano {\it et al.}~\cite{Ciriglianof+} & $0.984(12)$ & ChPT + large $N_c$ \\
Jamin, Oller, \& Pich~\cite{Jaminf+} & $0.974(11)$ & ChPT + dispersive (scalar form factor)  \\
Bijnens \& Talavera~\cite{BT03} & $0.976(10)$ & ChPT + Leutwyler \& Roos \\
\hline
Leutwyler \& Roos~\cite{LR84} & $0.961(8)$ & One-loop ChPT + quark model \\
\hline\hline
\end{tabular}
\end{table}

\subsection{\boldmath $\order(p^6)$ LEC combination $C_{12}^r+C_{34}^r$}
\label{sec:p6LECs}

The parameter $\tilde C_4$ in the two-loop ChPT fit function that we use to interpolate 
$f_+^{K\pi}(0)$ to the physical point---see Eq.~(\ref{eq:ChPTtwoloop})---is related to the 
combination of $\order(p^4)$ and $\order(p^6)$ LECs
\ba
\tilde C_4 = -\frac{8}{f_\pi^4}\left[C_{12}+C_{34}-L_5^2\right].
\ea
We can thus use the values of $\tilde C_4$ and $L_5^r$ from our fit output in 
Table~\ref{tab:priors} to extract the combination of $\order(p^6)$ LECs involved.
Taking correlations into account, we find 
\ba\label{eq:p6LECs}
\left[C_{12}^r+C_{34}^r\right](M_\rho) = 3.93(36)_\text{stat} (20)_\text{syst}\times10^{-6}\,. 
\ea
The first error in Eq.~(\ref{eq:p6LECs}) includes statistics, chiral
extrapolation and discretization errors, as well as the uncertainty from the LECs
(except $L_7$ and $L_8$) and the taste-violating hairpin parameters, as discussed
in Sec.~\ref{sec:systematic}. The second error is the sum in quadrature of the rest of the systematic
uncertainties. The detailed error budget is in Table~\ref{tab:LECserrors}. We obtain
all the errors in the same way as for $f_+^{K\pi}(0)$. Isospin corrections do not apply
to this quantity since it is defined in the isospin limit.
In practice, the values of LECs coming from a fit may be significantly affected by the presence or
absence of higher-order chiral terms in the fit function.  Therefore, applications of
our result in Eq.~(\ref{eq:p6LECs}) should allow the same type of corrections as in (the continuum limit of)  
\eq{ChPTtwoloop}. 
The complete error budget for this quantity can be found in Table~\ref{tab:LECserrors}.

Our result in Eq.~(\ref{eq:p6LECs}) agrees with nonlattice determinations in 
Refs.~\cite{Ciriglianof+,Jaminf+,LR84}. In those papers, the contribution to $f_+(0)$ from $C_{12}+C_{34}$ was calculated using the
  large $N_c$ approximation, a coupled-channel dispersion relation analysis, and a quark model,
  respectively.
  However, the value for  $C_{12}+C_{34}-L_5^2$ found in  Ref.~\cite{Kastner:2008ch}, which is
based on ChPT, large $N_c$ estimates of the LECs, and dispersive methods,
is $\sim3\sigma$ smaller than our value,
$\left[C_{12}^r+C_{34}^r-(L_5^r)^2\right](M_\rho) = \left(2.92\pm0.31\right)\times10^{-6}$. 

The result in Eq.~(\ref{eq:p6LECs}) also agrees very well with our previous calculation of this combination of LECs in
Ref.~\cite{Bazavov:2012cd}, on the MILC $N_f=2+1$ asqtad configurations, although with greatly reduced errors.
In fact, all sources of error are reduced due to several factors: the use of the MILC $N_f=2+1+1$ HISQ configurations with smaller
discretization errors than the asqtad action, data at smaller lattice spacings, data with physical light-quark masses, better tuning
of the strange sea quark masses, and including NLO finite-volume corrections explicitly.
The agreement with the JLQCD result in Ref.~\cite{Aoki:2017spo} is borderline, but the JLQCD calculation relies on simulations at a
single lattice spacing, although a systematic error is quoted for it, and it does not include data at the physical light-quark
masses.
Those systematics could affect more strongly the value of the combination of LECs than the form factor itself.

\begin{table}[tbp]
    \caption{Error budget for the LEC combinations of order $p^6$: $\left[C_{12}^r+C_{34}^r\right](M_\rho)$ and
    $\left[C_{12}^r+C_{34}^r-(L_5^r)^2\right](M_\rho)$.   \label{tab:LECserrors}}  
    \centering
\begin{tabular}{l@{\quad}c@{\quad}c}
   \hline\hline
    Source of uncertainty & $\left[C_{12}^r+C_{34}^r\right](M_\rho)\times 10^6$ &
        $\left[C_{12}^r+C_{34}^r-(L_5^r)^2\right](M_\rho)\times 10^6$ \\
    \hline
    Stat. + disc. + chiral inter. & $0.36$ & $0.23$ \\ 
    $L_{7,8}^r$ & $0.12$ & $0.13$ \\
    Scale $r_1$ & $0.13$ & $0.14$ \\ 
    $m_s^{\rm val}\ne m_s^{\rm sea}$ & $0.02$ & $0.02$  \\ 
    Finite volume & $0.09$ & $0.08$ \\ 
    \hline
    Total Error & $0.41$ & $0.031$ \\ 
    \hline\hline
\end{tabular}
\end{table}

%% file: phenomenology.tex
\section{Phenomenological implications}

\label{sec:pheno}

\subsection{Determination of $|V_{us}|$}

\label{sec:Vus}

Combining the form factor in \eq{fplusfinal} with the latest experimental average 
$|V_{us}|f_+^{K^0\pi^-}=0.21654(41)$ from Ref.~\cite{Moulson:2017ive}, we obtain
\ba \label{eq:ourVus}
|V_{us}| = 0.22333(44)_{f_+(0)}(42)_{{\rm exp}}= 0.22333(61)\,,
\ea
where the first error is from the uncertainty on the form factor, and the second is the experimental uncertainty.
Both errors are now of the same size.
The experimental error in Eq.~(\ref{eq:ourVus}) includes the uncertainty on the long-distance electromagnetic and strong
isospin-breaking corrections, $\delta_\text{EM}^{Kl}$ and $\delta_{{\rm SU(2)}}^{K\pi}$, which are taken into account when doing 
the experimental average of the neutral and charged modes~\cite{Moulson:2017ive}.
This uncertainty is however dominated by the errors in the lifetime and branching-ratio measurements of the neutral-kaon
modes~\cite{Moulson:2017ive}.
Other uncertainties such as those from the phase-space integrals are insignificant~\cite{Moulson:2017ive}.

\begin{figure}[tbp]
\centering
\includegraphics[width=1.\textwidth]{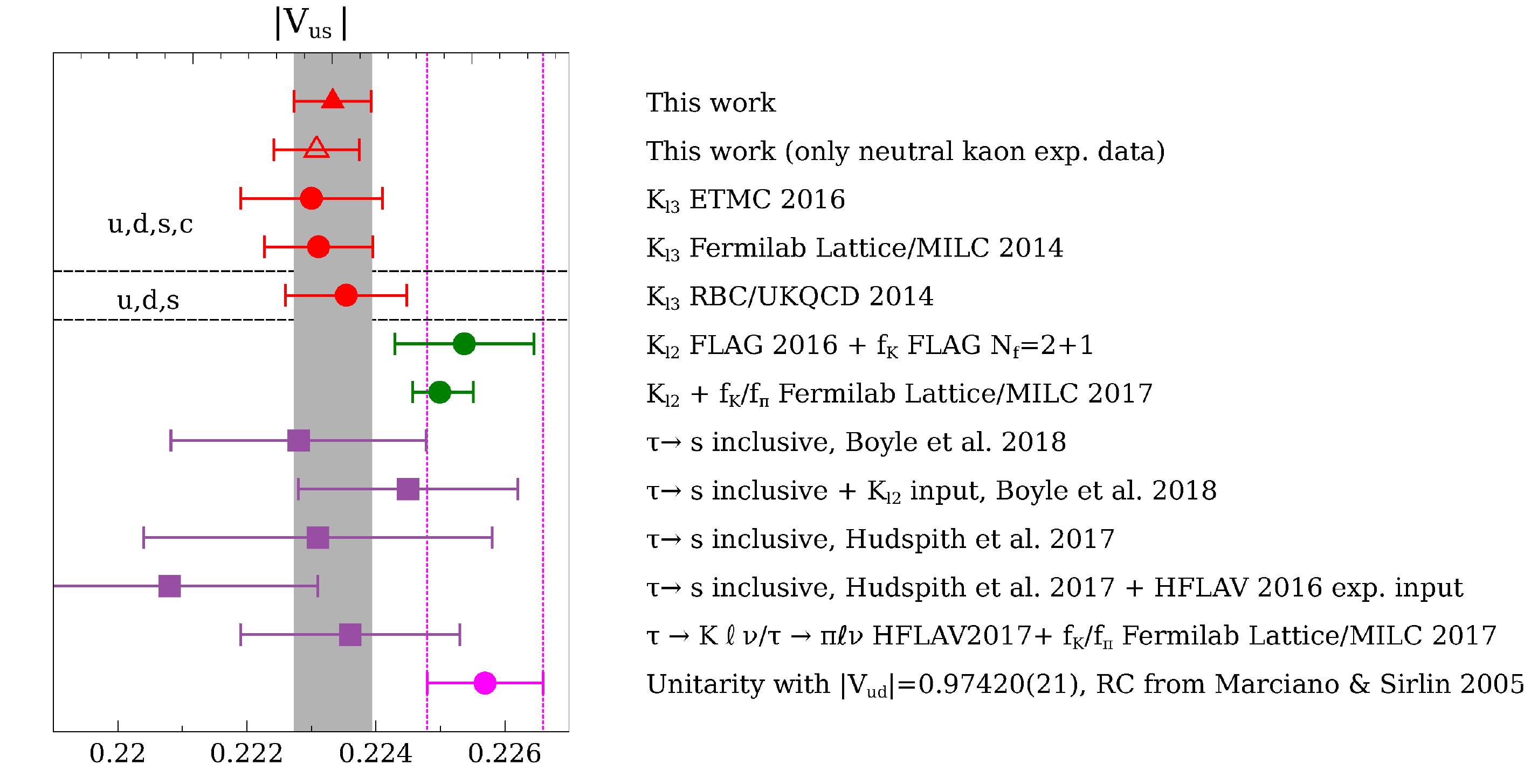}
\caption{Summary of recent $|V_{us}|$ determinations. 
The semileptonic determinations, labeled $K_{l3}$, use inputs for $f_+^{K\pi}(0)$ from the most recent lattice calculations in
    Refs.~\cite{Carrasco:2016kpy,Bazavov:2013maa,Boyle:2015hfa}, respectively.
    The leptonic determinations, labeled $K_{l2}$, use as inputs the $2+1$-flavor lattice-QCD average $f_K$ from 
    FLAG~\cite{Aoki:2016frl}, which only includes calculations where the lattice scale is set from physical inputs other than
    $f_\pi$, and the most recent and precise determination of $f_{K^\pm}/f_{\pi^\pm}$ from Ref.~\cite{Bazavov:2017lyh}.
    The inclusive hadronic $\tau$-decay determinations are the most recent ones, from Boyle et al.\ 2018~\cite{Boyle:2018ilm} and
    Hudspith et al.\ 2017~\cite{Hudspith:2017vew}.
    The second value from Ref.~\cite{Boyle:2018ilm} comes from relating the $\tau\to K\ell\nu$ branching fraction to the $K_{\mu 2}$
    branching fraction to get the experimental contribution from the $K$ pole. 
    The two values in Ref.~\cite{Hudspith:2017vew} correspond to using the normalization for $\tau$ decays into $K\pi$ modes as
    obtained in Ref.~\cite{Antonelli:2013usa} or as given by HFLAV~\cite{Amhis:2016xyh}.  
    For the exclusive $\tau$ determination we follow the calculation by the HFLAV group~\cite{Amhis:2016xyh}, but we update the value
    of the ratio $f_{K^\pm}/f_{\pi^\pm}$ to that in Ref.~\cite{Bazavov:2017lyh}.
    The unitarity value is taken to be $|V_{us}|=\sqrt{1-|V_{ud}|^2}$ with $|V_{ud}|$ from Ref.~\cite{Hardy:2018zsb}.
    RC stands for radiative corrections.
    The dotted magenta vertical lines correspond to this unitarity value.  
    The gray vertical band corresponds to our result in Eq.~(\ref{eq:ourVus}).
\label{fig:ResultsVus} }
\end{figure}

In Fig.~\ref{fig:ResultsVus} we compare our extraction of $|V_{us}|$ from $K$ semileptonic decays with other determinations using
$K$ semileptonic and leptonic decays, and hadronic $\tau$ decays.
Our semileptonic determination of $|V_{us}|$ is the most precise to date not relying on an external input for $|V_{ud}|$.
The central value agrees very well with the most recent lattice and nonlattice semileptonic calculations, as well as with those
based on hadronic tau decay, the latter have much larger errors.
Our result, however, is in tension with the leptonic determination using $f_K/f_\pi$ and with the unitarity prediction given by
$|V_{us}|=\sqrt{1-|V_{ud}|^2}$ with $|V_{ud}|$ from Ref.~\cite{Hardy:2018zsb}.
The agreement with the leptonic determination using $f_K$ is borderline.
The sizes of the disagreements---$2.6\sigma$ with unitarity and $2.2\sigma$ with the leptonic
determination using $f_{K}/f_{\pi}$---are similar to those using other recent lattice calculations for the semileptonic vector form
factor.

As a consistency check of the semileptonic extraction of $|V_{us}|$, we can consider 
the neutral- and charged-kaon modes separately.
Using our result in Eq.~(\ref{eq:fplusfinal}) together with the experimental average
for neutral modes only~\cite{Moulson:2017ive},
$|V_{us}|f_+^{K^0\pi^-}(0)=0.2163(5)$,\footnote{Notice that in order to perform
  the separate averages, Moulson~\cite{Moulson:2017ive} uses the phase-space
  integrals as extracted from the overall average of form-factor parameters.
  Although the phase-space factors are affected by isospin-breaking corrections,
  those corrections are expected to have a
  negligible impact at this level of precision since the uncertainty on the phase-space
  integrals currently has an insignificant impact on the experimental
  averages~\cite{Moulson:2017ive}.
}
we can compare $|V_{us}|$ as extracted exclusively from neutral-kaon decays: \\
$|V_{us}|_{K^0\pi^-}=0.22309(44)_{f_+(0)}(44)_{{\rm exp}}(25)_{\delta^{Kl}_{EM}}= 0.22309(67)$. 
In this case, we can disentangle the purely experimental error from the 
uncertainty in the long-distance electromagnetic corrections, 
$\delta_{EM}^{Kl}$, which is the same for all neutral modes,
  $\sim0.22\%$~\cite{Cirigliano:2004pv,Cirigliano:2001mk}. 
This result is in very good agreement with the value in
Eq.~(\ref{eq:ourVus}) within errors, which constitutes
a good test of the ChPT calculation of isospin (larger for the charged modes)
and EM (larger for the neutral modes) corrections included in the total experimental
average, as was already made clear by the results in Ref.~\cite{Moulson:2017ive}.

\subsection{Tests of CKM unitarity}

Using our main result for $|V_{us}|$ in Eq.~(\ref{eq:ourVus}), the value $|V_{ud}|=0.97420(21)$ from superallowed nuclear $\beta$
decays~\cite{Hardy:2018zsb}, and noting that $|V_{ub}|^2$ is negligible, we find that the measure
of deviation from first-row CKM unitarity in Eq.~(\ref{eq:unitarity}) is
\ba \label{eq:DeltaCKM}
\Delta_u\equiv\vert V_{ud}\vert^2+\vert V_{us}\vert^2+\vert           
V_{ub}\vert^2 -1 = -0.00104(27)_{V_{us}}(41)_{V_{ud}},
\ea 
which is $\sim 2.1\sigma$ 
away from the unitarity prediction, with an error dominated by the uncertainty on $\vert V_{ud}\vert$.
This makes revisiting the determination of $|V_{ud}|$ a priority for CKM tests. 
In this vein, one should examine not only superallowed $\beta$ decays but also other approaches.

At present, the precision in the extraction of $|V_{ud}|$ from the measurement of the neutron lifetime~\cite{Olive:2016xmw}
or pion $\beta$ decays~\cite{Pocanic:2003pf} is still far from that obtained from superallowed $\beta$ decays. 
In the case of superallowed $\beta$ decays, additional measurements will have a small effect on $|V_{ud}|$.
At the moment, the greatest improvement would come from a calculation of the short-distance radiative correction, which is 
the main source of uncertainty~\cite{Hardy:2018zsb}.
A~very recent calculation of the nucleus-independent contribution to those corrections, following a new methodology based on
dispersion relations~\cite{Seng:2018yzq}, obtains a value around $2\sigma$ larger than the current best determination by Marciano
and
Sirlin~\cite{Marciano:2005ec} and with a significant reduction of the error.
The increased electroweak radiative correction, when combined with the superallowed $\beta$ decay results~\cite{Hardy:2018zsb},
results in a lower value of~$|V_{ud}|$.
The authors of Ref.~\cite{Seng:2018yzq} quote $|V_{ud}|=0.97366(15)$.
Together with our result for $|V_{us}|$, this value of $|V_{ud}|$ considerably increases the tension with unitarity:
\ba 
\Delta_u\equiv\vert V_{ud}\vert^2+\vert V_{us}\vert^2+\vert           
V_{ub}\vert^2 -1 = -0.00209(27)_{V_{us}}(29)_{V_{ud}},
\ea 
a more than $5\sigma$ discrepancy.
We discuss further phenomenological implications of this new calculation in Sec.~\ref{sec:newVud}.
For the remainder of this section, we use the result by Marciano and Sirlin~\cite{Marciano:2005ec}, which leads to
$|V_{ud}|=0.97420(21)$ and Eq.~(\ref{eq:DeltaCKM}).

\begin{figure}[tbp]
    \centering
    \includegraphics[width=0.6\textwidth]{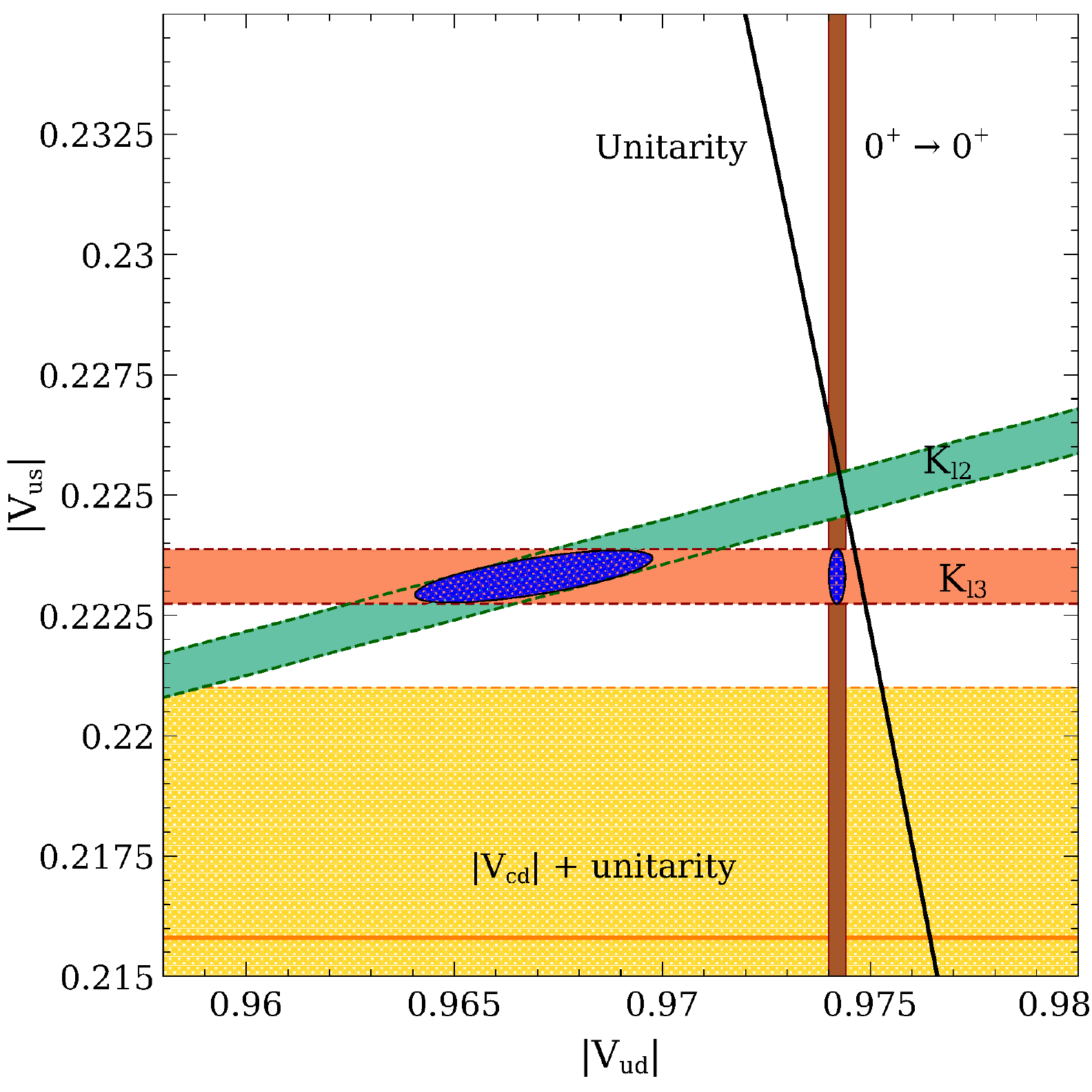}
    \caption{Constraints on $|V_{ud}|$ and $|V_{us}|$ from our results ($K_{l3}$), kaon leptonic decays ($K_{l2}$), superallowed
    nuclear $\beta$ decays, unitarity, and $|V_{cd}|$, as discussed in the text.
    Blue ellipses correspond to the allowed region from $K_{l3}$ and one of the other two constraints with a 68\% probability.
    Both regions have no overlap with unitarity (black line).
    Correlations between $K_{l2}$ and $K_{l3}$ are not taken into account.
    The orange horizontal line in the yellow region corresponds to the central value for $|V_{us}|$ as extracted from $|V_{cd}|$. 
    \label{fig:UTfirstrow} }
\end{figure}

To avoid using $|V_{ud}|$ as an input, we can instead perform a unitarity test relying only on experimental
kaon-decay measurements~\cite{Moulson:2017ive}, on the lattice input from the most recent
determination of $f_{K^+}/f_{\pi^+}$~\cite{Bazavov:2017lyh}, and on our result in
\eq{fplusfinal} for $f_+^{K^0\pi^-}(0)$. 
The result of the unitarity test using those inputs, noting again that $|V_{ub}|$ is negligible, 
is\footnote{The disentanglement of the EM and experimental errors in 
  Eq.~(\ref{eq:DeltaCKMlat}) is approximate, and intended only to indicate the relative
  size of these errors. The separation of the sources of error is 
  precise for leptonic decays, but for semileptonic decays we assume an overall
  $0.11\%$ EM error in the uncertainty of the experimental average. 
  This should be a fairly good approximation,  however, since the 
  average is dominated by the neutral modes for which the error is indeed $0.11\%$.}
\ba \label{eq:DeltaCKMlat}
\Delta_u\equiv\vert V_{ud}\vert^2+\vert V_{us}\vert^2+\vert V_{ub}\vert^2 -1 =
-0.0151(39)_{f_+(0)}(36)_{f_{K^\pm}/f_{\pi^\pm}}(36)_{{\rm exp}}(27)_\text{EM},
\ea
where the $2.2\sigma$ 
deviation from unitarity is a reflection of the tension between the leptonic and
semileptonic determinations of CKM matrix elements. These results are shown in 
Fig.~\ref{fig:UTfirstrow}, together with the test that takes  $|V_{ud}|$ from 
superallowed $\beta$ decays as input. No correlation between
$K_{l2}$ and $K_{l3}$ inputs, either on the theory or experimental sides, 
has been taken into account in this test.

One can perform another test of the unitarity of the CKM matrix by comparing 
$|V_{us}|$ with $|V_{cd}|$, which in the SM should be equal up to corrections of
$\order(\lambda^5)$, with $|V_{us}|=\lambda + \order(\lambda^7)$. Including  
  the ${\order(\lambda^5)}$ corrections, which only affect the last significant digit,
  the most precise determination of $|V_{cd}|=0.2151(6)_{f_D}(49)_\text{expt}(6)_\text{EM}$
  from leptonic decays~\cite{Bazavov:2017lyh} implies the value
  $|V_{us}|_{|V_{cd}|}=0.2158(52)$. This value of $|V_{us}|$ agrees at the $1.4\sigma$
  level with our result in Eq.~(\ref{eq:ourVus}), although
  with an uncertainty that is an order of magnitude larger. The uncertainty
  is dominated by the experimental error on the leptonic decay rate
  $D^+\to\ell^+\nu$, which is expected to be reduced by BESIII and Belle II. 
  This result is also depicted in Fig.~\ref{fig:UTfirstrow}.  
  As is the case for our main result, $|V_{us}|_{|V_{cd}|}$ is in tension with
  first-row CKM unitarity by about $2\sigma$ 
  when it is used together with $|V_{ud}|$ from superallowed nuclear $\beta$ decays 
  in Eq.~(\ref{eq:unitarity}).

Note that, in order to perform this test, we change the normalization of the decay constant $f_{D^+}$ obtained in
Ref.~\cite{Bazavov:2017lyh} to account for a change in the scale-setting quantity in that work, $f_{\pi^+}$, from the PDG value
$f_{\pi^+}=130.50\pm0.13~{\rm MeV}$~\cite{Olive:2016xmw} to the FLAG average $f_{\pi^+}=130.2\pm0.8~{\rm MeV}$~\cite{Aoki:2016frl}.
That gives us $f_{D^+}=212.2(0.3)_{{\rm stat}}(0.4)_{{\rm syst}}(1.2)_{f_\pi,\rm FLAG}[0.2]_\text{EM scheme}$.%
\footnote{Although the dependence of $f_{D^+}$ on the scale-setting quantity is much more complicated than a simple linear
relation, this estimate should capture most of the effect and, thus, be good
enough for this comparison, since its uncertainty is dominated by that of the
$D^+\to\ell^+\nu$ decay rate.}
The reason for that change is that the PDG value relies on an external input for $|V_{ud}|$, which is taken from superallowed
nuclear $\beta$ decays, which obscures the comparison.
The FLAG number, however, is an average of direct lattice determinations of $f_{\pi^+}$.
With this choice of $f_{\pi^+}$, the errors are fairly large, and the value of $|V_{ud}|$ extracted from experimental data on pion
leptonic decays~\cite{Rosner:2015wva} agrees within $\sim1.5\sigma$ with both $|V_{ud}|$ from superallowed nuclear $\beta$ decays
and the value from kaon decays only that we discuss below.

\subsection{Ratio of leptonic and semileptonic decays}

Another way of analyzing the tension between SM kaon leptonic and semileptonic 
decays is by looking at ratios of decay widths of leptonic and semileptonic 
decays, where the dependence on $|V_{us}|$ cancels. We can construct two ratios
\ba
\frac{\Gamma(K\to\ell\nu)}{\Gamma(K\to\pi\ell\nu)}\propto 
\left(\frac{f_{K^\pm}}{f_+^{K\pi}(0)}\right)^2,
\quad
\frac{\Gamma(K\to\ell\nu)/\Gamma(\pi\to\ell\nu)}{\Gamma(K\to\pi\ell\nu)}
\propto 
\frac{1}{|V_{ud}|^2}\left(\frac{f_{K^\pm}/f_{\pi^\pm}}{f_+^{K\pi}(0)}\right)^2.
\ea
The first ratio does not depend on any CKM matrix elements, while the second 
one is proportional to $1/|V_{ud}|^2$. In addition, the short-distance 
radiative corrections cancel between numerator and denominator in the first ratio,
but not in the second.

Taking experimental averages for the kaon decays and assuming the SM, we obtain%
\footnote{We take $\Gamma(K\to l\nu)$ from Ref.~\cite{Rosner:2015wva}, which does not use 
the same value of the universal short-distance electroweak correction $S_\text{EW}$ as \cite{Moulson:2017ive} (from which we take
the other
experimental averages). 
The imperfect cancellation is too small to affect the conclusion drawn here.} %
\cite{Rosner:2015wva,Moulson:2017ive}
\ba\label{eq:ratiosexp}
\left.\frac{f_{K^\pm}}{f_+^{K^0\pi^-}(0)}\right|_{{\rm exp.}}= 162.05(40)~{\rm MeV},\quad
\left.\frac{1}{|V_{ud}|}\frac{f_{K^\pm}/f_{\pi^\pm}}{f_+^{K^0\pi^-}(0)}\right|_{{\rm exp.}} = 1.2745(30).
\ea
With our result in \eq{fplusfinal} for $f_+^{K^0\pi^-}(0)$, the average of lattice calculations for
$f_{K^\pm}=155.6(0.4)~\text{MeV}$ from Ref.~\cite{Rosner:2015wva}, $f_{K^\pm}/f_{\pi^\pm}$ from Ref.~\cite{Bazavov:2017lyh}, and
$|V_{ud}|=0.97420(21)$ from \cite{Hardy:2018zsb}, those ratios are.
\ba\label{eq:ratioslatt}
\left.\frac{f_{K^\pm}}{f_+^{K^0\pi^-}(0)}\right|_\text{latt}= 160.58(79)~{\rm MeV},\quad
\left.\frac{1}{|V_{ud}|}\frac{f_{K^\pm}/f_{\pi^\pm}}{f_+^{K^0\pi^-}(0)}\right|_\text{latt} = 1.2651({}^{+31}_{-35}),
\ea
where we have not taken into account any correlation between the decay constants and the form factor.
Comparing Eqs.~(\ref{eq:ratiosexp}) and (\ref{eq:ratioslatt}), we see some tension, $\sim 1.7\sigma$ and $2.2\sigma$, 
respectively, between the SM predictions and the experimental measurements.
The error from lattice QCD is the main limiting factor in this comparison, but that can be reduced 
by taking into account the correlation between the numerator and denominator in
Eq.~(\ref{eq:ratioslatt}), which we plan to do in the future.

Alternatively, one can compare the ratio $\left[f_{K^\pm}/f_{\pi^\pm}\right]/\left[|V_{ud}|f_+^{K\pi}(0)\right]$ as extracted from
experiment and theory to get a value of the CKM matrix element $|V_{ud}|$, and compare it with the value from superallowed nuclear
$\beta$ decays.
The result of such an exercise is
$|V_{ud}|=0.9669(19)_{f_+}({}^{+13}_{-19})_{f_K/f_\pi}(23)_\text{exp}=0.9669({}^{+32}_{-35})$, approximately 
$2.1\sigma$ lower than the value from superallowed $\beta$ decays. 
This result is seen in Fig.~\ref{fig:UTfirstrow} at the intersection of the two bands for $K_{\ell3}$ and
$K_{\ell2}$. It also deviates from the unitarity condition.

The unitarity test comparing $|V_{us}|/|V_{ud}|$ with $|V_{us}|_{|V_{cd}|}/|V_{ud}|$, again including corrections up to
${\order(\lambda^5)}$, and taking the decay constants $f_{K^+}/f_{\pi^+}$ and $f_{D^+}/f_{\pi^+}$ from Ref.~\cite{Bazavov:2017lyh}
and the experimental data on leptonic experimental data from Ref.~\cite{Rosner:2015wva}, fails at the $2\sigma$ level.
This test is limited by the experimental error on the $D^+$ leptonic decay rate.

\subsection{\boldmath Implications of the new extraction of $|V_{ud}|$}
\label{sec:newVud}

  If the decrease of the central value and uncertainty of the nucleus-independent
  electroweak radiative
  corrections involved in the extraction of $|V_{ud}|$ from superallowed $\beta$ decays
  in Ref.~\cite{Seng:2018yzq} is confirmed, the new value $|V_{ud}|=0.97366(15)$
  would exacerbate some of the tensions we have just discussed.

  First, as shown above, this value of $|V_{ud}|$ and our semileptonic result
  for $|V_{us}|$ would imply a greater than $5\sigma$ violation of first-row CKM unitarity.
  The tension between our semileptonic value of $|V_{us}|$ and the one extracted from
  kaon leptonic decays and $f_{K^\pm}/f_{\pi^\pm}$, however, would be slightly reduced to
  $2\sigma$, since a smaller value of $|V_{ud}|$ would give a smaller value of
  the leptonic $|V_{us}|$, closer to our semileptonic extraction. For the same reason,
  the tension between the ratios involving $|V_{ud}|$ in Eqs.~(\ref{eq:ratiosexp}) and
  (\ref{eq:ratioslatt}) would be slightly lessened.

\begin{figure}[tbp]
    \centering
    \includegraphics[width=1.\textwidth]{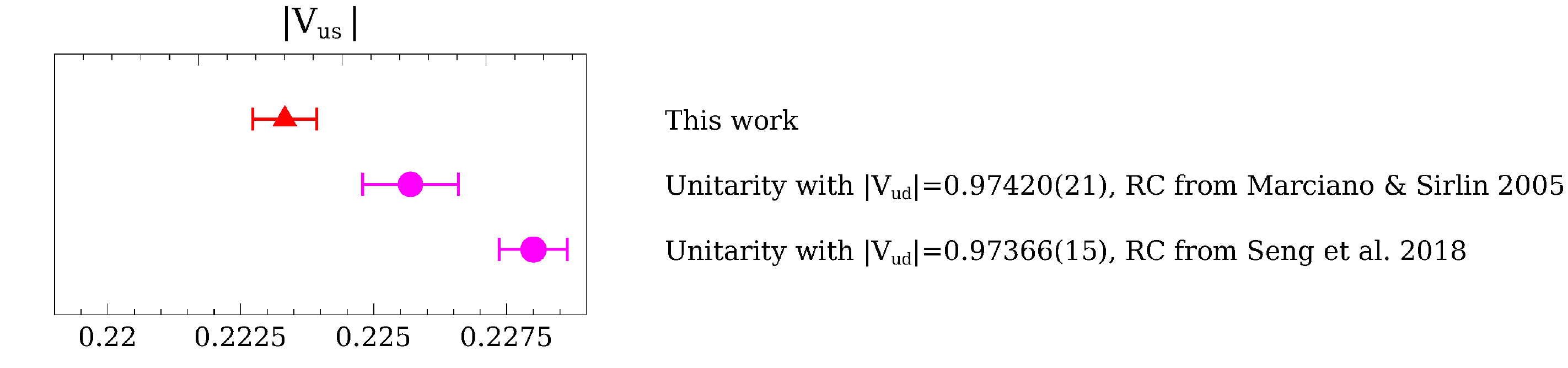}
    \caption{Comparison of the unitarity point using $|V_{ud}|=0.97366(15)$ with the results in this work, and with the unitarity
      point corresponding to $|V_{ud}|=0.97420(21)$.
      RC stands for radiative corrections.
        \label{fig:ResultsVusnewVud} }
\end{figure}

In Fig.~\ref{fig:ResultsVusnewVud}, as an example, we show the comparison of the unitarity prediction $\sqrt{1-|V_{ud}|^2}$ for
$|V_{us}|$ using both $|V_{ud}|=0.97366(15)$ and $|V_{ud}|=0.97420(21)$, together with the results in this work.
Given the important implications of a value of $|V_{ud}|$ with a smaller error and a smaller central value, it is very important to
confirm the new calculation of radiative corrections in Ref.~\cite{Seng:2018yzq}, and to understand the discrepancy with the
previous best determination in Ref.~\cite{Marciano:2005ec}.

\section{Conclusions and outlook}
\label{sec:conclusions}

Using the HISQ $N_f=2+1+1$ MILC ensembles, we have performed the most precise computation
to date of the vector form factor at zero momentum transfer, $f_+^{K^0\pi^-}(0)$, and the
first one to include the dominant FV effects, as calculated in ChPT at NLO.  Our result
for the form factor enables a direct determination of the CKM matrix element $|V_{us}|$
from semileptonic kaon decays with a theory error that is, for the first time, at the
same level as the experimental error. 
Further,  the uncertainty in this direct determination is now similar to those
from indirect determinations based on leptonic decays with $|V_{ud}|$ as input.

A key to achieving  this level of precision is simulating at near-physical values
of the quark masses, which drastically reduces the systematic errors associated with the chiral 
extrapolation (replacing it with an interpolation), as well as the error coming from
the chiral LECs that are inputs to the analysis. The finite-volume effects, one of
the main sources of uncertainty in our previous analyses, have also been significantly
reduced by explicitly including them at NLO (the leading nontrivial order) in ChPT.
The dominant remaining source of error is now statistics, 
which could be reduced by extending
the key ensembles with physical quark masses, and including the existing MILC
physical-mass ensemble with a finer lattice spacing of $a\approx0.042$~fm.

Another important error arises from the uncertainty in the ChPT LECs of order~$p^6$.
That uncertainty could be reduced by performing a combined analysis of form-factor data together with light meson masses and decay
constants, which would put more constraints on the ChPT LECs.
In particular, the error from $L_8$ is comparable to, but greater than, that from $L_7$, and the combined analysis could
significantly reduce the $L_8$ error.
Errors from $L_4$, $L_5$, and $L_6$ would also be reduced, but they have a much smaller effect on the total error here.

We find that the extraction of $|V_{us}|$ from semileptonic kaon decays is in tension both with the extraction from leptonic kaon
decays and with unitarity at the $\sim 2$--$2.6\sigma$ level.
In particular, the unitarity test based only on kaon decay data, without any external input for $|V_{ud}|$, and having as
nonperturbative inputs $f_+^{K^0\pi^-}(0)$ from this work and $f_{K^\pm}/f_{\pi^\pm}$ from Ref.~\cite{Bazavov:2017lyh}, shows a 
$\sim 2.2\sigma$ tension.
While unitarity tests based on $|V_{ud}|$ are currently limited by the uncertainty in that matrix element, the tension with
unitarity would raise to the $5\sigma$ level if the new calculation of radiative corrections involved in the extraction of
$|V_{ud}|$ from superallowed $\beta$ decays~\cite{Seng:2018yzq} is confirmed.

The test based on kaon-decay data has similarly sized uncertainties arising from both theory and experiment.
In order to shed light on these tensions, improvements from both the theoretical and experimental
sides are urgently needed, as are improvements in other approaches.
A new round of experiments is expected to reduce the experimental error to $\sim 0.12\%$ in the next few
years~\cite{Moulson:2017ive}.
More importantly, the new high-statistics data will help to check the consistency of current fits, and to perform a more thorough
study of systematic errors on the experimental averages.

For the experimental determination of $|V_{us}|f_+^{K^0\pi^-}$, electromagnetic and isospin effects are currently being estimated
using phenomenology and ChPT techniques.
Although they are not yet a dominant source of error (EM effects make a 0.11\% correction to the individual neutral channels), with
the reduction of other sources of error and the forthcoming improvement in the experimentally measured branching ratios and
lifetimes, they will eventually need to be included directly in the lattice-QCD simulations.
Recent efforts in that direction can be found in Refs.~\cite{Carrasco:2015xwa,Giusti:2017dwk,Patella:2017fgk,SachrajdaLat18}.

Isospin corrections are numerically important for the charged kaon channels, where those 
effects enter already at LO through $\pi^0$-$\eta$ mixing. The NNLO ChPT estimate of 
the corrections for the charged modes has large errors~\cite{Bijnens:2007xa} due to the
unknown value of the $\order(p^6)$ LECs. Fortunately, the experimental average is dominated
by the neutral-kaon channels, so the charged-mode uncertainty
does not have a large effect on the final
experimental average. The strong isospin-breaking correction $\delta_\text{SU(2)}$ used in the
experimental average is a NLO ChPT estimate that partially includes NNLO effects;
it does not include the uncertainty associated with higher-order terms in the
chiral expansion. However, the fact that the value 
used in the average and the one extracted from experiment are so close (2.45(19)\% vs.\
2.82(38)\%~\cite{Moulson:2017ive}), that the result for $|V_{us}|$ using only the neutral
modes agrees with the one using all decay modes---see Sec.~\ref{sec:Vus}, and 
that neutral modes are the dominant ones in the average, indicates that the experimental
average using this estimate is robust.

The uncertainties from the phase-space integrals are insignificant at present in the 
final error for the experimental average. It is therefore not crucial at present to
have a better representation of those, i.e., to have the $q^2$ dependence of the form
factors. In the future, however, lattice calculations of $f_+^{K\pi}(q^2)$ could provide
better determinations of the form factor slope than those relying on experimental
data~\cite{Carrasco:2016kpy,Moulson:2017ive}.

An important future step in the investigation of the tensions observed in the first-row unitarity relation, and in the value of
$|V_{us}|$ extracted from different sources, will be to perform a correlated analysis of semileptonic and leptonic kaon decays.
That analysis would provide a more precise value of the ratio
$\left[f_{K^\pm}/f_{\pi^\pm}\right]/\left[|V_{ud}|f_+^{K\pi}(0)\right]$ and potentially give an insight into the tensions.
Another key point in the study of those tensions is clarifying the role of the electroweak radiative corrections in the extraction
of $|V_{ud}|$ from superallowed $\beta$ decays, as well as reducing the error of that CKM matrix element as extracted, not only from
superallowed $\beta$ decays, but from other sources.